\documentstyle[12pt]{article}
\makeatletter \oddsidemargin  -.1in \evensidemargin -.1in
\newcommand{\singlespacing}{\let\CS=\@currsize\renewcommand{\baselinestretch}{1}\tiny\CS}
\newcommand{\oneandahalfspacing}{\let\CS=\@currsize\renewcommand{\baselinestretch}{1.25}\tiny\CS}
\newcommand{\doublespacing}{\let\CS=\@currsize\renewcommand{\baselinestretch}{1.35}\tiny\CS}

\newtheorem{rule-def}[theorem]{Rule}

\RequirePackage[dvips]{graphicx} \textheight 22.5cm

\begin{document} 

\title {\bf Peristaltic Transport of a Rheological Fluid: Model for Movement of Food Bolus Through Esophagus}
\author{\small J.C.Misra$^1$\thanks{Email address: {\it misrajc@rediffmail.com (J.C.Misra)}}, ~~~S. Maiti$^2$\thanks{Email address: {\it somnathm@cts.iitkgp.ernet.in (S.Maiti)}}~  \\
\it$^1$Department of Mathematics,\\ Institute of Technical Education
and Research,\\Siksha O Anusandhan University, Bhubaneswar-751030, India\\
$^2$\it School of Medical Science and Technology $\&$ Center for
Theoretical Studies, \\Indian Institute of Technology, Kharagpur-721302, India \\}
\date{}
\maketitle \noindent \doublespacing

\begin{abstract}
Fluid mechanical peristaltic transport through esophagus has been of
concern in the paper. A mathematical model has been developed with an
aim to study the peristaltic transport of a rheological fluid for
arbitrary wave shapes and tube lengths. The Ostwald-de Waele power law
of viscous fluid is considered here to depict the non-Newtonian
behaviour of the fluid. The model is formulated and analyzed with the
specific aim of exploring some important information concerning the
movement of food bolus through the esophagus. The analysis has been
carried out by using lubrication theory. The study is particularly
suitable for cases where the Reynolds number is small. The esophagus
is treated as a circular tube through which the transport of food
bolus takes places by periodic contraction of the esophageal
wall. Variation of different variables concerned with the transport
phenomena such as pressure, flow velocity, particle trajectory and
reflux are investigated for a single wave as well as for a train of
periodic peristaltic waves.  Locally variable pressure is seen to be
highly sensitive to the flow index `n'. The study clearly shows that
continuous fluid transport for Newtonian/rheological fluids by wave
train propagation is much more effective than widely spaced single
wave propagation in the case of peristaltic movement of food bolus in
the esophagus.  
\end{abstract} 

\section{Introduction}

Swallowing of food is a mechanical process that begins with chewing,
smashing and mixing of food in the oral cavity. Complex structural
motion is set in within the pharynx that forces the food bolus rapidly
into the esophagus. The process ends with the movement of the bolus
into the stomach by peristaltic contraction of the esophageal
wall. Pumping through various vessels of the physiological system by
means of propagation of peristaltic waves is considered by
physiologists as a natural mechanism of pumping materials in the case
of most fluids of the physiological system.  Besides physiological
applications, the benefit of studies on peristaltic movement, however,
extends to a variety of industrial appliances, e.g. roller pumps used
to pump caustic or corroding liquids. Many of the essential fluid
mechanical characteristics of peristalsis have found important
applications in different engineering problems investigated by several
researchers. Studies on peristalsis have also many important
applications in the design and construction of many useful devices of
biomedical engineering and technology, such as artificial blood
devices, for example, finger pumps used in the pumping of blood. Our
earlier communications (Misra et al. \cite{Misra1,Misra2,Misra3},
Maiti and Misra \cite{Maiti}) and also those of some other authors
\cite{Guyton,Jaffrin1,Nadeem,Hayat} provide useful information
regarding peristaltic transport of various types of fluids.
\begin{center}
\begin{tabular}{|l l|}\hline
{~\bf Nomenclature} &~ \\
~~$R,\theta,Z$ & Cylindrical co-ordinates\\
~~$a$ & Average radius of the food bolus\\
~~$H$ & Displacement of the esophageal wall in the radial direction\\
~~$n$ & Fluid index number\\
~~$k$ & Reciprocal of n\\
~~$P$ & Fluid pressure\\
~~$Q_1$ & Volume flow rate\\
~~$t$ & Time\\
~~$V_B$ & Volume of fluid within a single peristaltic wave (the
bolus)\\
~~$U,V,W$ & Velocity components in Z-, R-, $\theta$- directions respectively\\
~~$\delta $ & wave number\\
~~$\Delta P$ & Pressure difference between the ends of the esophagus\\
~~$\epsilon$ & Minimum vessel radius (during occlusion)\\
~~$\lambda$ & Wave length of the travelling wave motion in the esophagus\\
~~$\mu $ & Viscosity of the fluid (food bolus)\\
~~$\nu $ &  Kinematic viscosity of the fluid (food bolus)\\
~~$\phi$ Wave amplitude\\
~~$\rho$ & Fluid density\\
\hline
\end{tabular}
\end{center}
Solid/liquid food mixture or chyme transport through esophagus which
is a muscular conduit leading to the stomach takes place by means of
progression of peristaltic contraction waves of circular muscle fibers
contracted within circular muscle layers of the esophageal wall. When
peristaltic waves start propagating, the circular muscle cells shorten
themselves causing contractile forces.  Involvement of both the nerve
control and the intrinsic properties of muscle cells complicates the
mechanism of muscle contraction. Consequently the peristaltic
contraction acts as an external force on the tissue structure and
travels downwards with a certain speed. The length of the esophagus is
250-300 mm for an adult human being. When stretched, it becomes more
or less a straight tube that extends between the pharynx and the
stomach. The two ends of the esophagus are controlled by the upper
esophageal sphincters (UES) and the lower esophageal sphincters
(LES). During resting condition, a high contractile pressure of at
least 30 mm Hg is maintained. The intraluminal pressure at rest above
the UES is maintained equal to the atmospheric pressure. In the thorax
the luminal pressure at rest is typically slightly below the
atmospheric pressure, while in the abdomen the pressure is about 10 mm
Hg above the atmospheric pressure. The thoracic as well as the
intra-abdominal pressures are adjusted by respiration and is
maintained at about 5 mm Hg. During the pharyngeal phase of
swallowing, a mass of food that has been chewed at the point of
swallowing, called bolus passes rapidly through the pharynx. Thereby
the UES relaxes to atmospheric pressure, and the bolus arrives at the
esophagus. As the intra-bolus pressure adds to about 5 mm Hg, a
peristaltic contraction wave passes through the UES and then
progresses down the esophagus at a rate of 20-40 mm/s, transporting
the fluid bolus distally.  Following the initiation of swallowing, the
LES actively relaxes in a while to gastric pressure and discloses as
the esophageal peristaltic wave forces the bolus into the stomach
\cite{Brasseur}. Esophageal peristalsis acts as a pump in transporting
a fluid bolus from the upper esophagus to the stomach.  Total pumping
is not achieved, if the esophagus fails to maintain complete
occlusion. Often in the region where the aortic arch impresses upon
the esophagus, the esophageal wall occlusion remains incomplete. As a
result, some fluid bolus leaks proximally through the contracted
region and is left behind. When LES that helps keep the acidic
contents of the stomach out of the throat does not work properly,
laryngopharyngeal reflux occurs.

This leads to various discomforts of the body. For example, an
individual may feel bitter test in the throat, uneasiness in
swallowing of food bolus, feel burning sensation/pain in the throat
and other similar health problems related to stomach. It may be
mentioned that most of the studies on peristaltic transport made by
previous authors are not very suitable for applications to those
physiological situations where a single wave travels down the length
of an organ having finite dimensions (e.g. esophagus). Li and
Brasseur \cite{Li1} dwelt on the said aspects of peristaltic
pumping. They presented a model of peristaltic transport of a
Newtonian viscous fluid, for arbitrary wave shapes/arbitrary wave
number through a finite length tube. The conventional sinusoidal
wave equation was developed by considering the position of the wall
as a function of the minimum radius of the tube, which vibrates only
in one direction. This study  has got limited application. It is
applicable only when the intake is water or some drink having
similar physical properties.

But the movement of food grain bolus, like whipped cream, custard,
ketchup, suspensions of corn starch and various masticated
food-grains through the esophageal tube exhibits non-Newtonian
behaviour. It is, therefore, important to study the peristaltic
transport of the food bolus, when the motion is predominantly
non-Newtonian. While studying the rheological behaviour of some
physiological fluids, Patel et al. \cite{Patel} carried out
experimental investigation and reported some data for some
biorheological fluids. These data indicated that the masticated
food-grains may be treated as a power-law fluid, where the power-law
index may vary, depending on the type of the food material.

Keeping this in view, a mathematical model has been developed here to
study the peristaltic transport of food bolus through the esophagus,
by considering the motion to be governed by Ostwald-de Waele power law
\cite{Bird}. The fluid transport by peristalsis has been approximated
by the lubrication theory which holds for Re$\le 1$
(cf. \cite{Jaffrin2}). It may be mentioned that that when the average
bolus is almost equal to the wave length ($\delta\sim 1$), the
lubrication theory gives a reasonably good approximation to the
pressure field (cf. \cite{Dusey}).  The present study is undertaken to
address the basic fluid mechanical issue of the non-steady effects
corresponding to the finite tube length in the case of rheological
(non-Newtonian) fluids. Our prime concern has been to examine the
difference of the magnitudes of the flow variables in the cases of
Newtonian and rheological (non-Newtonian) fluids. Particular emphasis
has been paid to investigate the variation of essential local variable
pressure together with volume flow rate, the pressure difference
between the ends of the tube, representing the esophagus, the velocity
distribution, the particle trajectories and the reflux
phenomenon. Based upon the present study, a useful comparison has been
made between the single wave and wave train effects on the peristaltic
transport characteristics of the movement of food bolus.

\section{Mathematical Modelling}
In studies pertaining to friction dominated flows where axial length
scale of velocity variation is large in comparison to the radial
scales, use of the lubrication theory has been found to be very
effective. It is known that transport of food bolus through esophagus
takes place by the mechanism of peristalsis, where viscosity crosses
the threshold limit 200 cp.  The peristaltic wave speed c
(characteristic velocity) in this case is normally 20-40 mm$/$s so
that Reynolds number is of order 0.001-1. Let us treat the esophagus
as an axi-symmetric tube of length L (which usually ranges between 250
mm and 300 mm) and denote by $\epsilon$ the minimum tube radius (i.e
tube occlusion) and the wave number by $\delta=a/\lambda$. The ratio
between average bolus radius $a=(V_B/\pi \lambda)^\frac{1}{2}$ (5-10
mm) and the typical wave length $\lambda$ (50-100 mm) is of order
0.05-0.2, where $V_B$ stands for the fluid volume within a single
peristaltic wave (bolus).  \\
\begin{figure}
 \includegraphics[width=6.0in,height=4.0in]{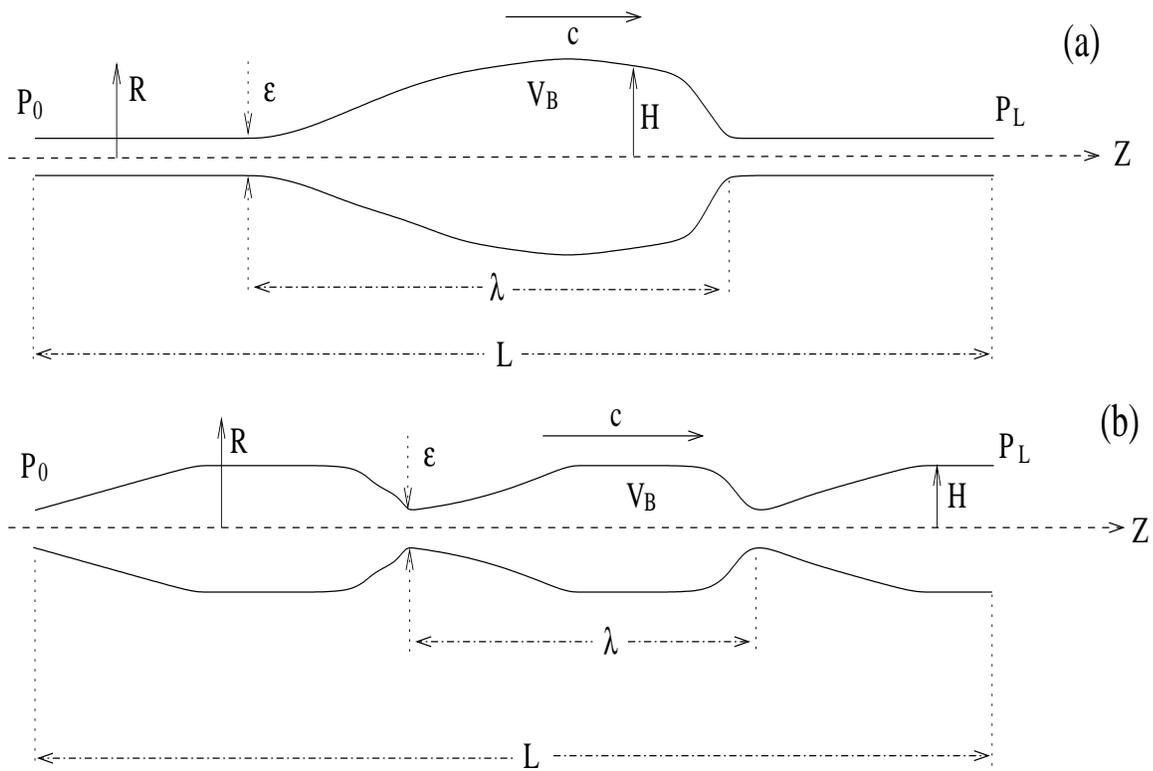}
\caption{Schematic diagram of the problem$:$ (a) a single
contraction wave, (b) wave train. In both the cases, the food bolus
is supposed to move from left to right against a pressure difference
$P_L-P_0$, by peristaltic contraction waves along the tube.  A
non-integral number of peristaltic waves in the tube (L/$\lambda$)
is depicted in (b).} \label{paper6_geo6.1.1}
\end{figure}
We take (R,$\theta$,Z) as the cylindrical coordinates of some
location of a fluid particle, R being the radius of the tube; Z is
measured in the direction of wave propagation. The schematic diagram
of peristaltic transport has been given in Fig.
\ref{paper6_geo6.1.1} that illustrates (a) a single wave moving
along a finite tube and (b) continual production of multiple waves.
Let R=H(Z,t) denote the shape of the esophageal wall.

The mathematical model developed here pertains to a situation, where
the fluid mechanical peristaltic transport of a food bolus is driven
by arbitrarily shaped deformation of the wall of the esophagus. The
pressure boundary conditions at the ends of the esophagus will also
be considered, when its length is taken to be finite. The food bolus
will be treated as an incompressible viscous Ostwald-de Waele type
of rheological fluid \cite{Bird}. If $\tau$ be the stress tensor and
$\Delta$ the symmetric rate of deformation tensor, the constitutive
equation for the fluid can be written as
\begin{equation}
\tau=\alpha\left\{|\sqrt{\frac{1}{2}(\Delta:\Delta)}|^{n-1}\right\}\Delta,
\end{equation}
\begin{eqnarray*}
where~~\frac{1}{2}(\Delta:\Delta)=2\left(\left(\frac{\partial V}{\partial R}\right)^2+\left(\frac{V}{R}\right)^2+\left(\frac{\partial U}{\partial Z}\right)^2\right)+\left(\frac{\partial U}{\partial R}+\frac{\partial V}{\partial R}\right)^2
\end{eqnarray*}
 in which $\alpha$ and n denote respectively the consistency factor and
 the power law index parameter, depicting the behaviour
 of the fluid. It is known that a shear thinning fluid
 is characterized by $n<1$, while for a shear thickening fluid, $n>1$. Based on the above consideration, the motion
 of food bolus in the esophagus can be considered to be governed by the
 equations

\begin{equation}
\rho \left (\frac{\partial U}{\partial t}+U\frac{\partial
U}{\partial Z}+V\frac{\partial U}{\partial R}\right
)=-\frac{\partial P}{\partial Z}+\frac{1}{R}\frac{\partial
  (R\tau_{RZ})}{\partial R}+\frac{\partial \tau_{ZZ}}{\partial Z}
\end{equation}
\begin{equation}
\rho\left (\frac{\partial V}{\partial t}+U\frac{\partial V}{\partial
Z}+V\frac{\partial V}{\partial R}\right )=-\frac{\partial
P}{\partial R}+\frac{1}{R}\frac{\partial
  (R\tau_{RR})}{\partial R}+\frac{\partial \tau_{RZ}}{\partial Z}
\end{equation}

\section{Analysis}

In the model, each material point on the wall of the esophagus is
considered to move in the radial direction with velocity $\partial
H(Z,t)/\partial t$. The following
non-dimensional variables will be introduced in the analysis that
follows:
\begin{eqnarray}
&\bar{Z}&=\frac{Z}{\lambda}, ~~\bar{R}=\frac{R}{a},
~~\bar{U}=\frac{U}{c}, ~~\bar{V}=\frac{V}{c\delta}, ~~
\delta=\frac{a}{\lambda},
  ~~\bar{P}=\frac{a^{n+1}P}{\alpha c^n\lambda},\bar{Q}=\frac{\eta Q_1}{\pi a^2c}~~\nonumber
 \\ &\bar{t}&=\frac{ct}{\lambda}, ~\bar{H}=\frac{H}{a},
 ~ ~~ Re=\frac{\rho a^n}{\alpha c^{n-2}},
 \bar{\tau_0}=\frac{\tau_0}{\alpha(\frac{c}{a})^n}, \bar{\tau_{RZ}}=\frac{\tau_{RZ}}{\alpha (\frac{c}{a})^n},
  ~~\nonumber\\
\end{eqnarray}
where $\eta=1$ for wave train and $\eta=L/\lambda$ for single wave movement.
In terms of these variables the governing equations can be rewritten
as (dropping the bars over the symbols)
\begin{equation}
 Re\delta \left (\frac{\partial U}{\partial t}+U\frac{\partial
U}{\partial Z}+V\frac{\partial U}{\partial R}\right
)=-\frac{\partial P}{\partial Z}+\frac{1}{R}\frac{\partial
  \left(\Phi\left(R\frac{\partial U}{\partial R}+R\delta^2\frac{\partial V}{\partial
Z}\right)\right)}{\partial R}+2\delta^2\frac{\partial \left(\Phi\frac{\partial
U}{\partial Z}\right)}{\partial Z}
\end{equation}
\begin{equation}
Re\delta^3\left (\frac{\partial V}{\partial t}+U\frac{\partial V}{\partial
Z}+V\frac{\partial V}{\partial R}\right )=-\frac{\partial
P}{\partial R}+\delta^2\frac{1}{R}\frac{\partial
  (R\Phi\frac{\partial V}{\partial R})}{\partial R}+\delta^2\frac{\partial\left(\Phi(\frac{\partial U}{\partial R}+\delta^2\frac{\partial V}{\partial
Z})\right)}{\partial Z}
\end{equation}
\begin{eqnarray}
\Phi=\left|\sqrt{2\delta^2\left\{\left(\frac{\partial V}{\partial R}\right)^2+\left(\frac{V}{R}\right)^2+\left(\frac{\partial
U}{\partial Z}\right)^2\right\}+\left(\frac{\partial U}{\partial R}+\delta^2\frac{\partial V}{\partial
Z}\right)^2}\right|^{n-1}
\end{eqnarray}
Considering the wall curvature as very small $(\delta \ll 1 )$, it is
possible to apply the lubrication theory, where the inertial effect is
negligible and the dominant radial scale 'a' is quite small, in comparison
to the dominant axial scale $\lambda$. In such a case, the distribution of pressure is uniform on
each cross section. Under these considerations, the governing equations and the boundary
conditions in terms of non-dimensional variables reduce to the
following set of equations :
\begin{equation}
0=-\frac{\partial P}{\partial Z}+ \frac{1}{R}\frac{\partial
  (R\frac{\partial U}{\partial R}|\frac{\partial U}{\partial
    R}|^{n-1})}{\partial R}
\label{paper6_zmomentum_lubrication}
\end{equation}
\begin{equation}
0=-\frac{\partial P}{\partial R}
\end{equation}

\begin{eqnarray}
\frac{\partial U}{\partial R}=0,~ V=0~
at~R=0;~~U=0~,~V=\frac{\partial H}{\partial t} at~ R=H
\label{paper6_boundary_condition_1}
\\P=P_0~at~Z=0~and~P=P_L~at~Z=L
\label{paper6_boundary_condition_2}
\end{eqnarray}
By solving (\ref{paper6_zmomentum_lubrication}) subject to the
conditions (\ref{paper6_boundary_condition_1}) and
(\ref{paper6_boundary_condition_2}), we find the velocity field in
the form
\begin{equation}
U(R,Z,t)=\frac{p|p|^{k-1}}{2^k(k+1)}\left[R^{k+1}-H^{k+1}\right]
\label{paper6_axial_velocity}
\end{equation}
\begin{equation}
V(R,Z,t)=\frac{Rp|p|^{k-1}}{2^k(k+1)}\left[p_1\left(\frac{H^{k+1}}{2}-\frac{R^{k+1}}{k+3}\right)+
\frac{k+1}{2}H^k \frac{\partial H}{\partial Z}\right]
\label{paper6_tranverse_velocity}
\end{equation}
where~$p=\frac{\partial P}{\partial Z},~k=\frac{1}{n}$ and
$p_1=k\frac{\partial p}{\partial Z}/p$.  \\Now using the last of the
conditions (\ref{paper6_boundary_condition_1}), we have from
(\ref{paper6_tranverse_velocity}) the equation
 \begin{eqnarray}
\frac{\partial H}{\partial
t}=\frac{H^{k+1}p|p|^{k-1}}{2^{k+1}(k+3)}\left[p_1H+(k+3)\frac{\partial
H}{\partial Z}\right]
\label{paper6_H_relation_pressure_gradient_pressure}
\end{eqnarray}
The pressure gradient p obtained on integrating
(\ref{paper6_H_relation_pressure_gradient_pressure}) is given by
\begin{eqnarray}
p|p|^{k-1}=\frac{2^{k+1}(k+3)}{H^{k+3}}[c_1+\int_{0}^{Z}H\frac{\partial
H}{\partial t}dZ], \label{paper6_pressure_gradient}
\end{eqnarray}
where $c_1$ is, in general, a function of time t. Solving
(\ref{paper6_pressure_gradient}), we obtain
\begin{eqnarray}
P(Z,t)-P(0,t)=\int_{0}^{Z}p(S,t)dS~~~~~~~~~~~~~~~~~~~~~~~~~~~~~~~~~~~~~~~~~~~~~~~~~~~~~~~
\nonumber\\=\int_{0}^{Z}\left[\left|\frac{2^{k+1}(k+3)}{H^{k+3}}
\left\{c_1+\int_{0}^{S}H\frac{\partial H}{\partial
t}dZ\right\}\right|^{n-1}\left\{\frac{2^{k+1}(k+3)}{H^{k+3}}
\left\{c_1+\int_{0}^{S}H\frac{\partial H}{\partial
t}dZ\right\}\right\}\right]dS \label{paper6_pressure_rise}
\end{eqnarray}
Using (\ref{paper6_axial_velocity}) along with
(\ref{paper6_pressure_gradient}), the non-dimensionalized volume
flow rate is given by
\begin{eqnarray}
\bar{Q}(Z,t)=2\eta\int_{0}^{H}RUdR\nonumber~~~\\=-\frac{\eta
p|p|^{k-1}H^{k+3}}{2^k(k+3)}
\label{paper6_volume_flow_rate_related_pressure}
\\=-2\eta\left\{c_1+\int_{0}^{Z}H\frac{\partial H}{\partial t}dZ\right\}
\label{paper6_volume_flow_rate_without_pressure_term}
\end{eqnarray}
Putting Z=0, the instantaneous flow rate at the inlet of the esophagus
is given by
\begin{eqnarray}
\bar{Q}(0,t)=-2\eta c_1 \label{paper6_volume_flow_rate_inlet}
\end{eqnarray}
 In terms of the flow rate $\bar{Q}$(0,t)
 at the inlet, the temporal flow rate $\bar{Q}(Z,t)$  at any position of the
 esophagus can be expressed as
\begin{eqnarray}
\bar{Q}(Z,t)=\bar{Q}(0,t)-2\eta \int_{0}^{Z}H\frac{\partial
H}{\partial t}dZ
\label{paper6_volume_flow_rate_interms_inlet_flow_rate}
\end{eqnarray}
Using (\ref{paper6_pressure_rise}),
(\ref{paper6_volume_flow_rate_inlet}) and
(\ref{paper6_volume_flow_rate_interms_inlet_flow_rate}), the flow
rate $\bar{Q}(Z,t)$ is found to be related to the pressure P(Z,t) as
\begin{eqnarray}
P(Z,t)-P(0,t)=-\int_{0}^{Z}\left|\frac{2^k(k+3)\bar{Q}(Z,t)}{\eta
H^{k+3}}\right|^{(1/k)-1}\left\{\frac{2^k(k+3)\bar{Q}(Z,t)}{\eta
H^{k+3}}\right\}dZ
\label{paper6_pressure_rise_related_volume_flow_rate}
\end{eqnarray}
Thus the pressure difference between the esophageal ends is given by
\begin{eqnarray}
\Delta P=P(L,
t)-P(0,t)\nonumber\\=-\int_{0}^{L}\left|\frac{2^k(k+3)\bar{Q}(Z,t)}{\eta
H^{k+3}}\right|^{(1/k)-1}\left\{\frac{2^k(k+3)\bar{Q}(Z,t)}{\eta
H^{k+3}}\right\}dZ
\end{eqnarray}
It is worthwhile to note that the equation
(\ref{paper6_volume_flow_rate_related_pressure})  reduces to the
corresponding equation derived by Li and Brasseur \cite{Li1} who
studied a similar problem for a Newtonian fluid.
\section{Numerical Study}

\begin{figure}
\includegraphics[width=3.5in,height=2.1in]{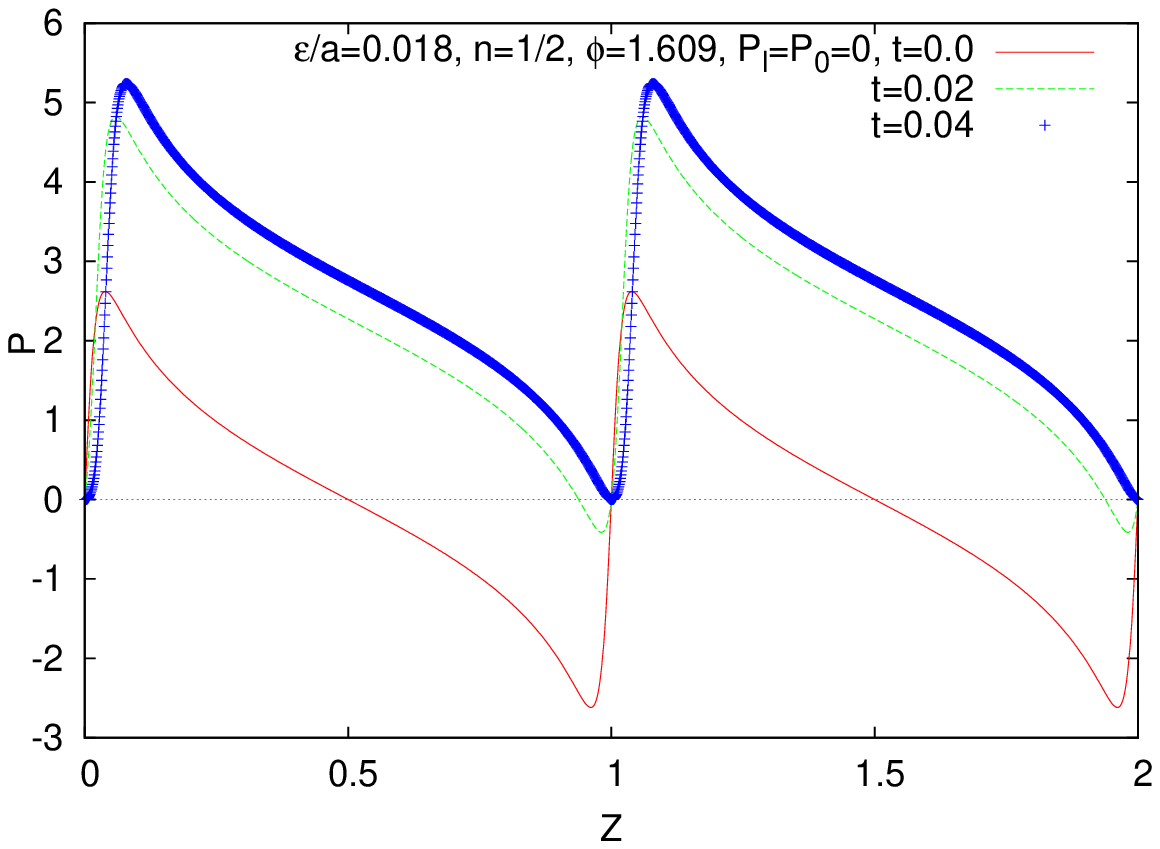}\includegraphics[width=3.5in,height=2.1in]{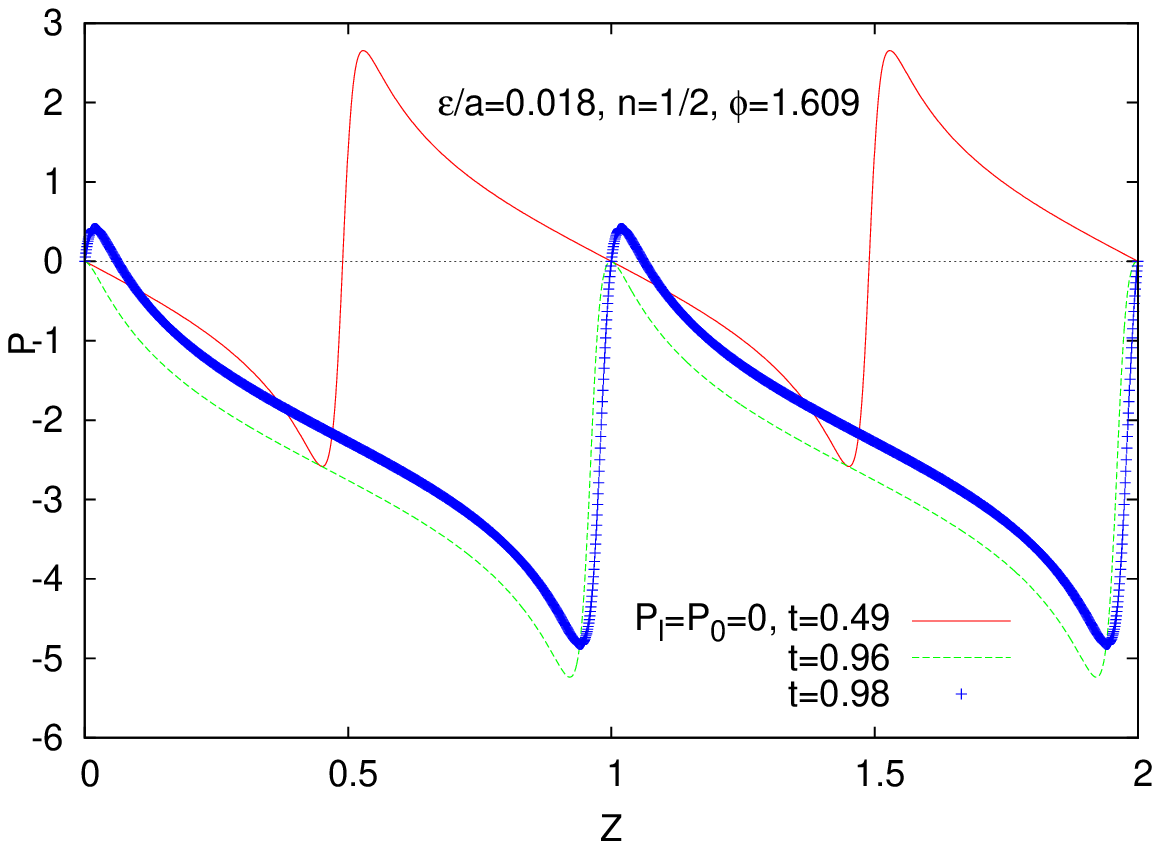}
\\$~~~~~~~~~~~~~~~~~~~~~~~~~~~(a)~~~~~~~~~~~~~~~~~~~~~~~~~~~~~~~~~~~~~~~~~~~~~~~~~~~~~~~~~~~~~~~~~~~~~(b)~~~~~~~~~~~~~~~$\\
\includegraphics[width=3.5in,height=2.1in]{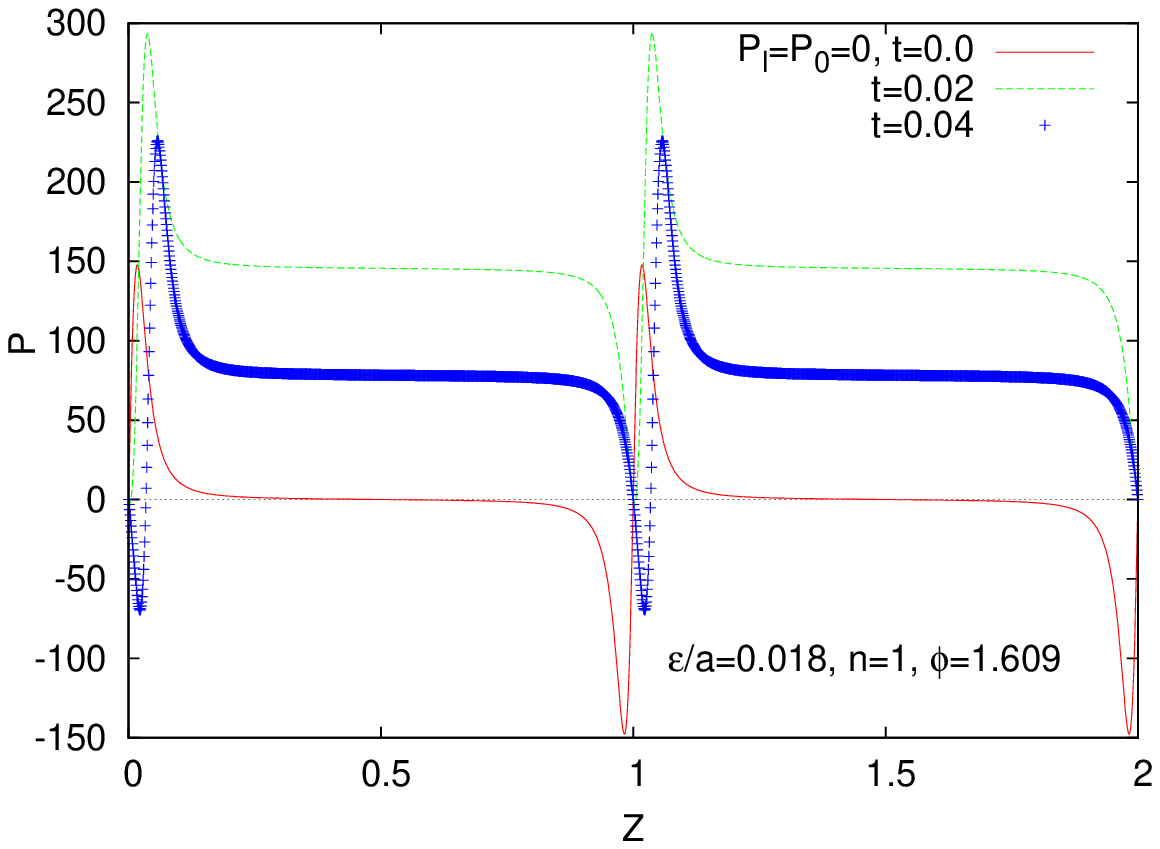}\includegraphics[width=3.5in,height=2.1in]{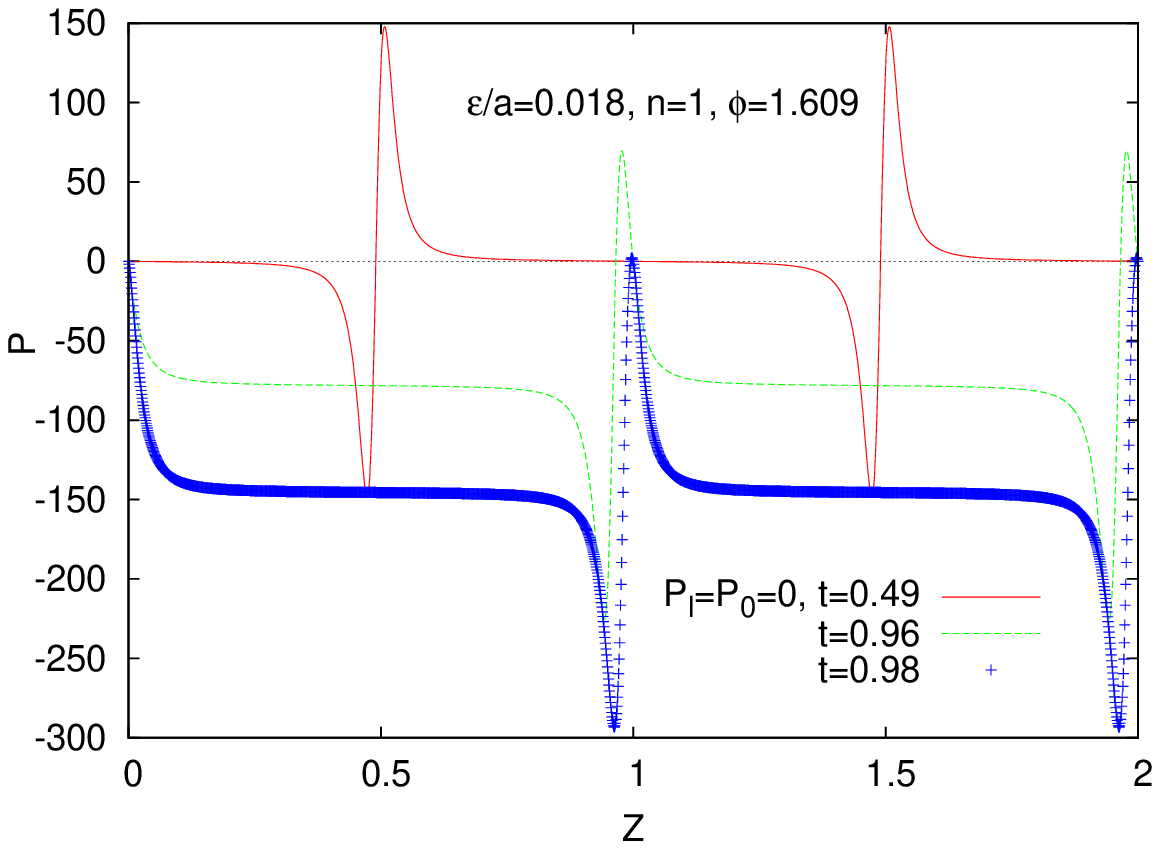}
\\$~~~~~~~~~~~~~~~~~~~~~~~~~~~(c)~~~~~~~~~~~~~~~~~~~~~~~~~~~~~~~~~~~~~~~~~~~~~~~~~~~~~~~~~~~~~~~~~~~~~(d)~~~~~~~~~~~~~~~$\\
\includegraphics[width=3.4in,height=2.1in]{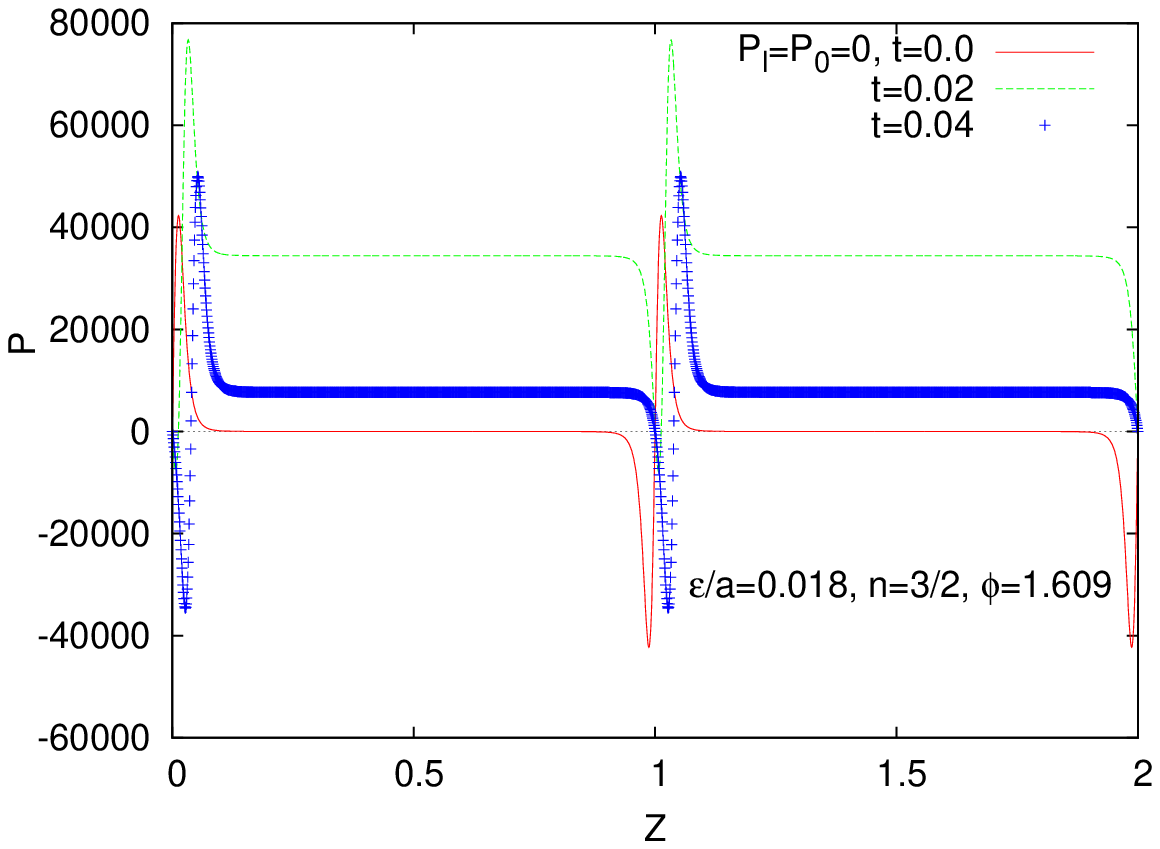}\includegraphics[width=3.5in,height=2.1in]{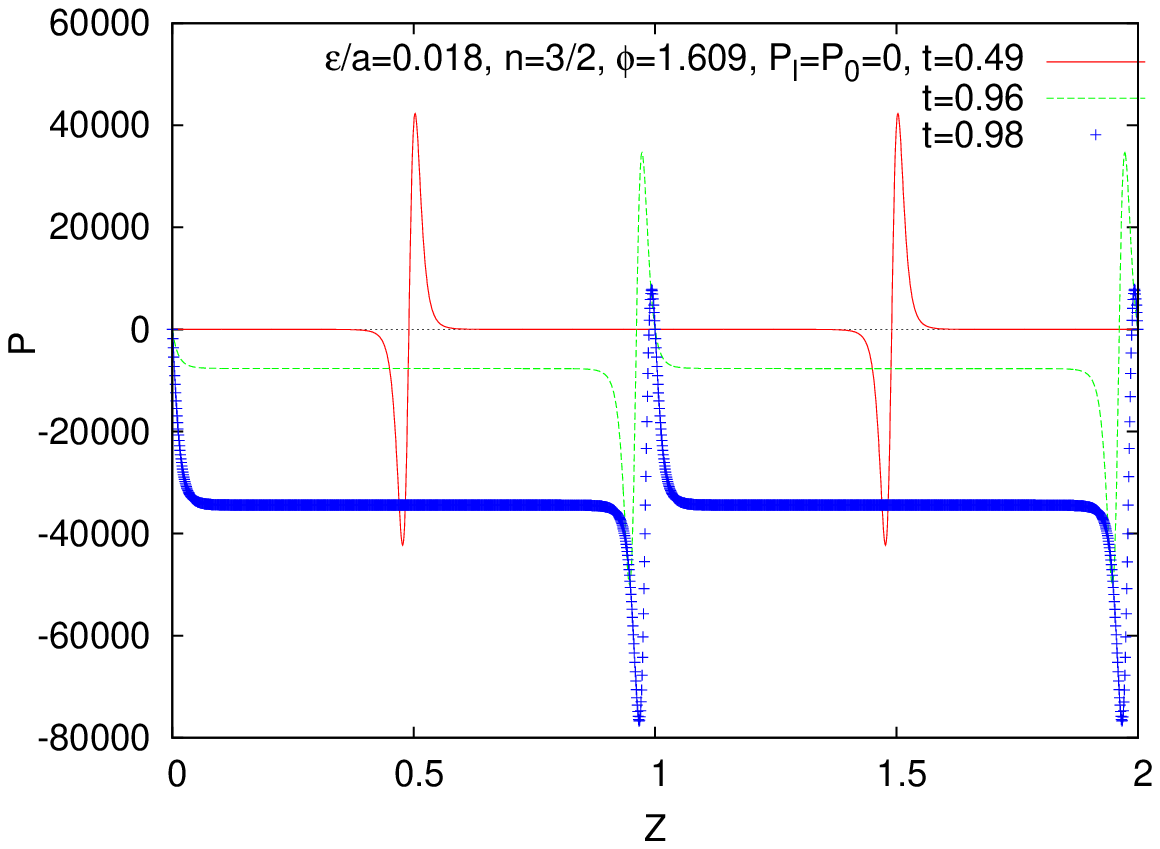}
\\$~~~~~~~~~~~~~~~~~~~~~~~~~~~(e)~~~~~~~~~~~~~~~~~~~~~~~~~~~~~~~~~~~~~~~~~~~~~~~~~~~~~~~~~~~~~~~~~~~~~(f)~~~~~~~~~~~~~~~$\\
\caption{Local Pressure Distribution in the case of a wave train in
the esophagus. These figures reveal that for a shear thinning fluid the (global) maximum and minimum peaks are attained respectively at t=0.4 and t=0.96. This is contrary to the cases of Newtonian and shear thickening fluids for which the maximum and minimum values are attained at t=0.02 and t=0.98 respectively.}
\label{paper6_pressurec6.1.1.1-6.1.3.4}
\end{figure}
This section is devoted to a quantitative analysis of the
mathematical model presented in the earlier sections. We shall try
to investigate the difference between the characteristics in the
cases of single wave and multiple wave (i.e. wave train) for the
peristaltic transport of food bolus for arbitrary wave shapes and
esophageal length. It may be noted that unlike in the study of Li
and Brasseur \cite{Li1} (for the Newtonian case), it is not possible
to find a closed form solution for $c_1$ that appears in
(\ref{paper6_pressure_gradient}), (\ref{paper6_pressure_rise}) and
(\ref{paper6_volume_flow_rate_without_pressure_term}). Consequently
it is not possible to find an explicit analytical expression for the
fluid flux $\bar{Q}$ when the tube length and wave shape are both
arbitrary. Determination of the quantitative estimates of different
physical variables has been based upon the consideration that for
the rheological (non-Newtonian) fluid taken up in our present study
the flow rate $\bar{Q}(Z,t)$ is given by
\begin{eqnarray}
\bar{Q}^n(Z,t)=Q^n+H^2-\frac{1}{\eta}\int_{0}^{\eta}H^2dt,
\label{paper6_volume_flow_assumed_form}
\end{eqnarray}
Q being the time-averaged volume flow and the superscript `n'
denoting the power law index of the fluid.

\subsection{Pressure Distribution}
Let us first investigate the effect of finite tube length on the
pressure distribution during the peristaltic transport. It may be
noted that pressure is essentially a mechanical variable in the
functioning of the esophagus where intraluminal manometry is used as
a common diagnostic tool in order to obtain the contractile
characteristics of the circular muscle within the esophageal wall.
Let us first take up the case of an integral number of train waves
moving with constant speed through a tube having finite length whose
ends are subjected to constant pressure of equal magnitude. During
peristalsis, the esophagus is considered to be of sinusoidal shape
defined by the equation
\begin{eqnarray}
H(Z,t)=\epsilon/a+0.5\phi \{1-\cos 2\pi(Z-t)\}
\end{eqnarray}

\begin{figure}
\includegraphics[width=3.5in,height=2.0in]{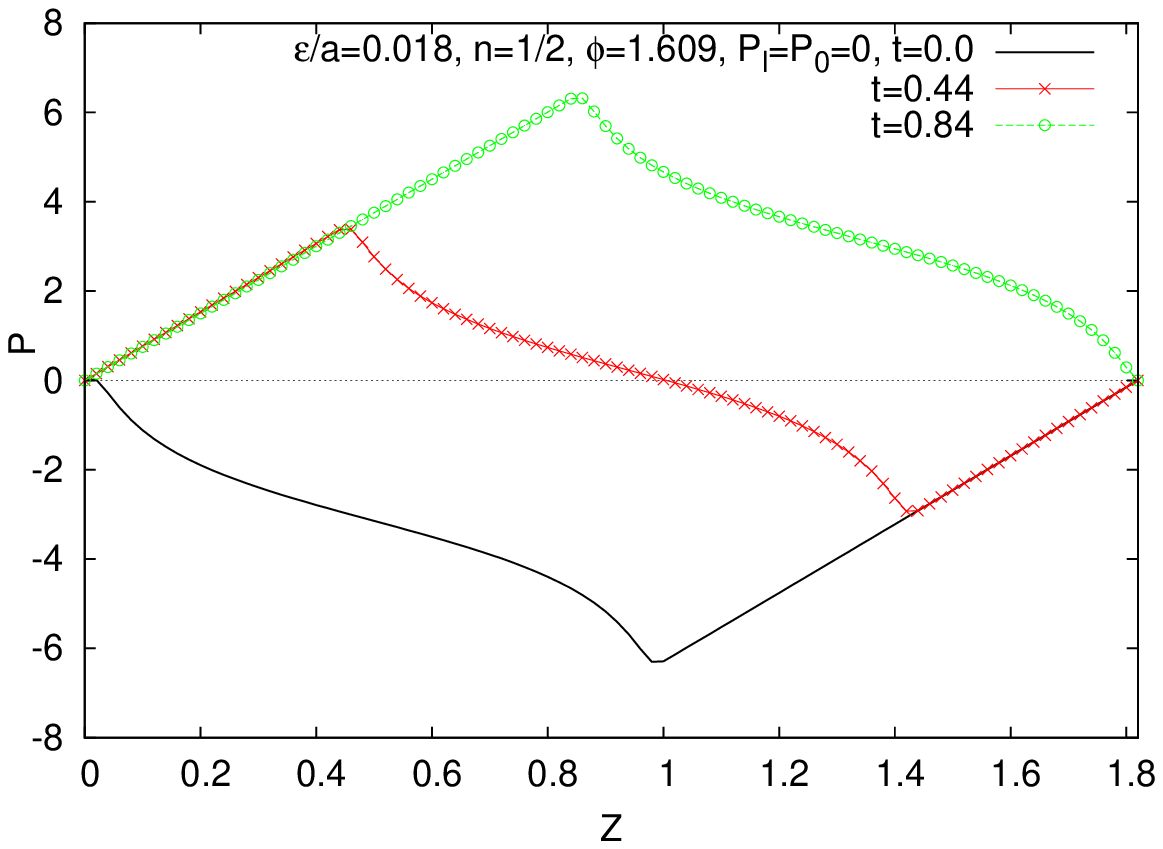}\includegraphics[width=3.5in,height=2.0in]{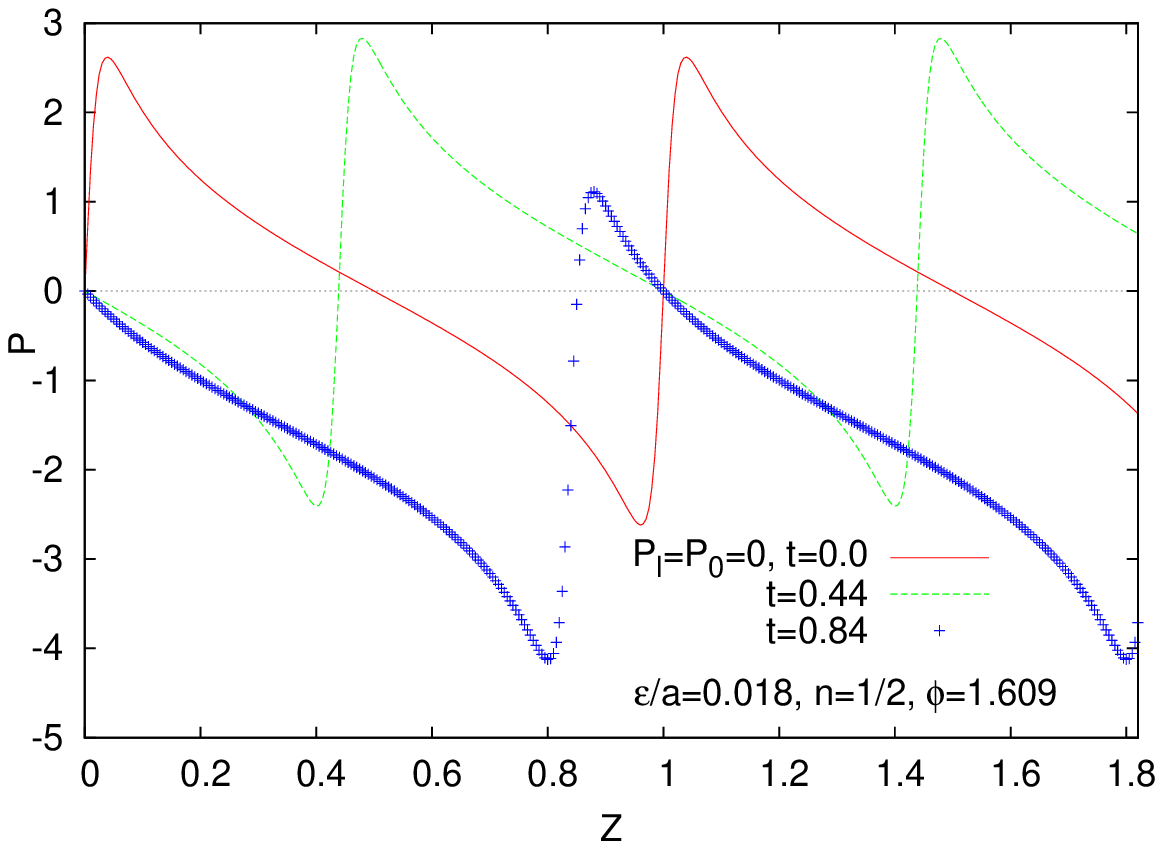}
\\$~~~~~~~~~~~~~~~~~~~~~~~~~~~(a)~~~~~~~~~~~~~~~~~~~~~~~~~~~~~~~~~~~~~~~~~~~~~~~~~~~~~~~~~~~~~~~~~~~~~(b)~~~~~~~~~~~~~~~$\\
\includegraphics[width=3.5in,height=2.0in]{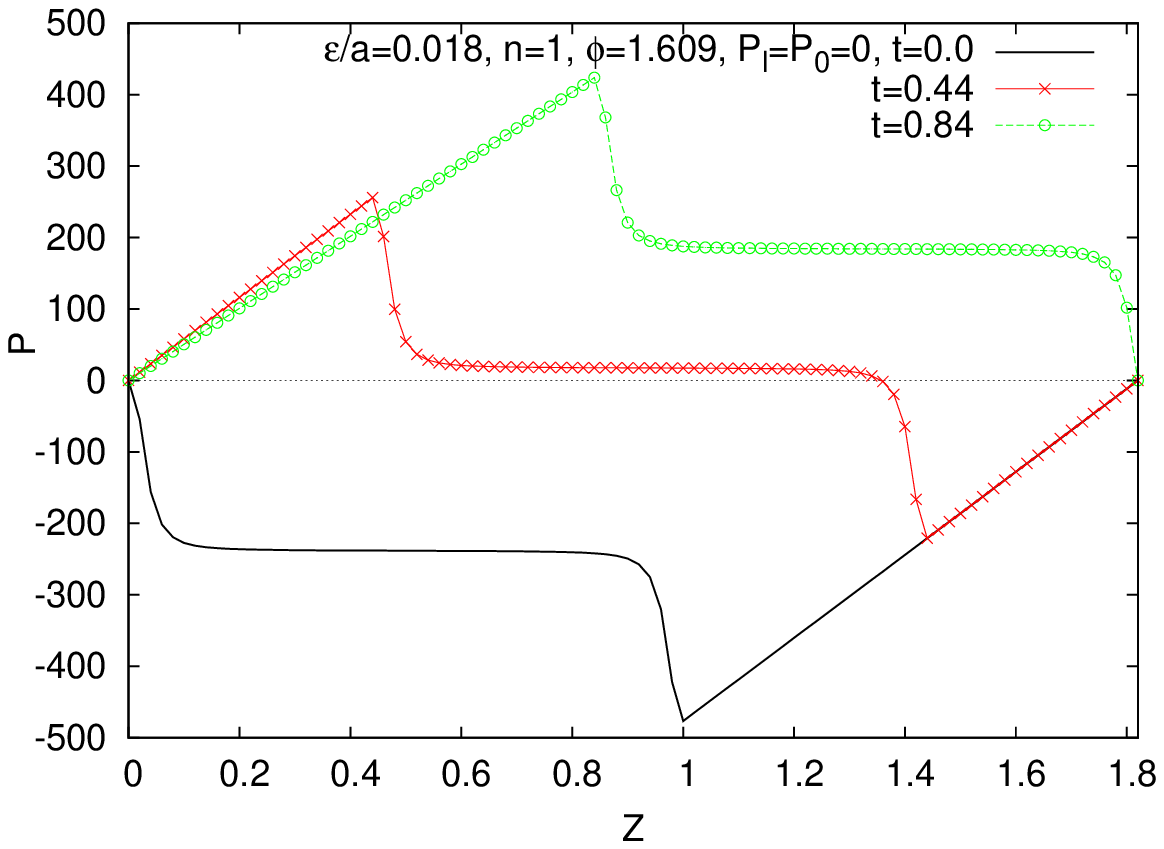}\includegraphics[width=3.5in,height=2.0in]{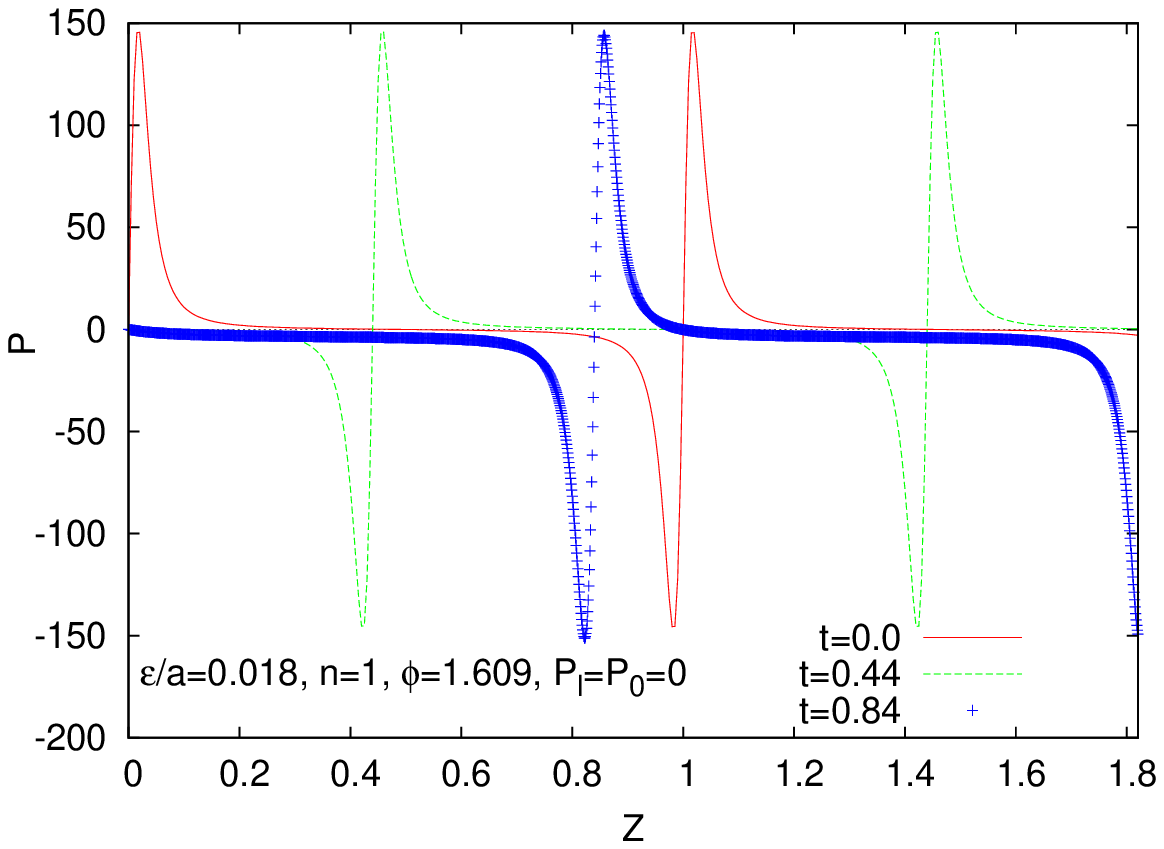}
\\$~~~~~~~~~~~~~~~~~~~~~~~~~~~(c)~~~~~~~~~~~~~~~~~~~~~~~~~~~~~~~~~~~~~~~~~~~~~~~~~~~~~~~~~~~~~~~~~~~~~(d)~~~~~~~~~~~~~~~$\\
\includegraphics[width=3.4in,height=2.0in]{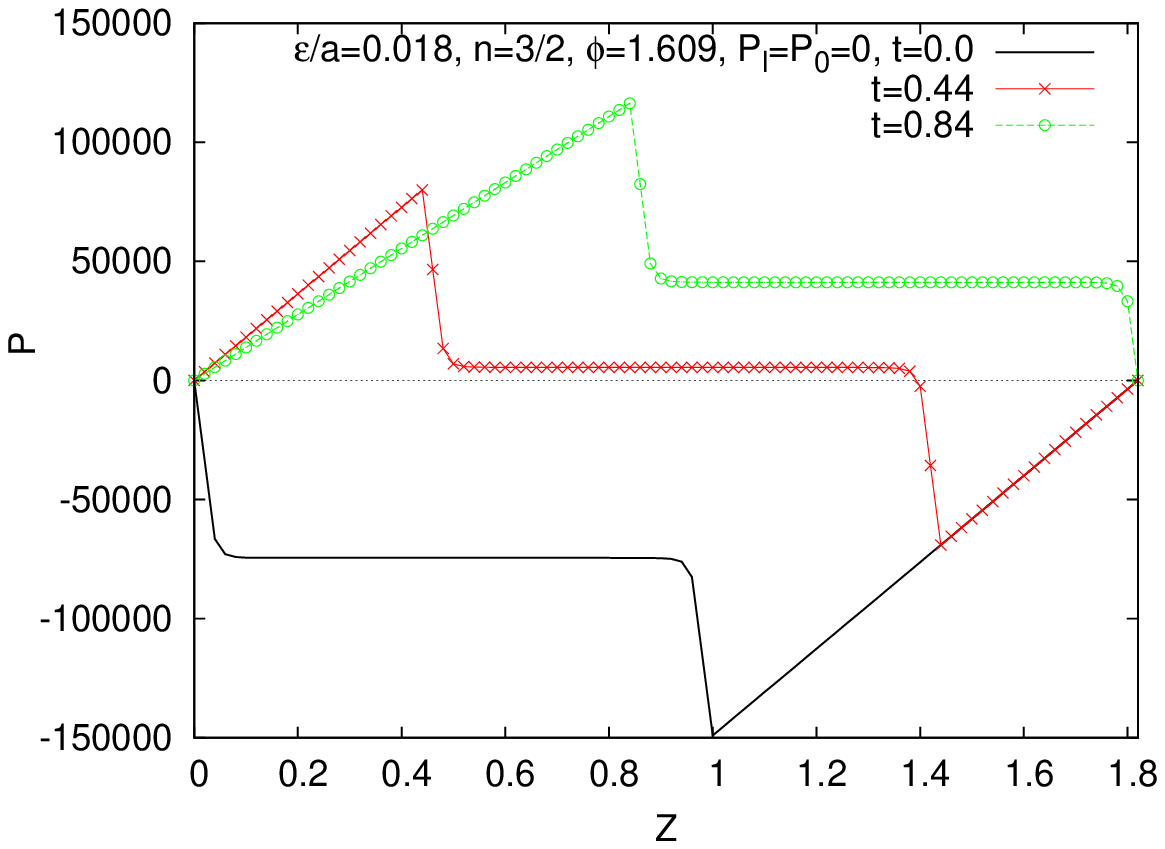}\includegraphics[width=3.5in,height=2.0in]{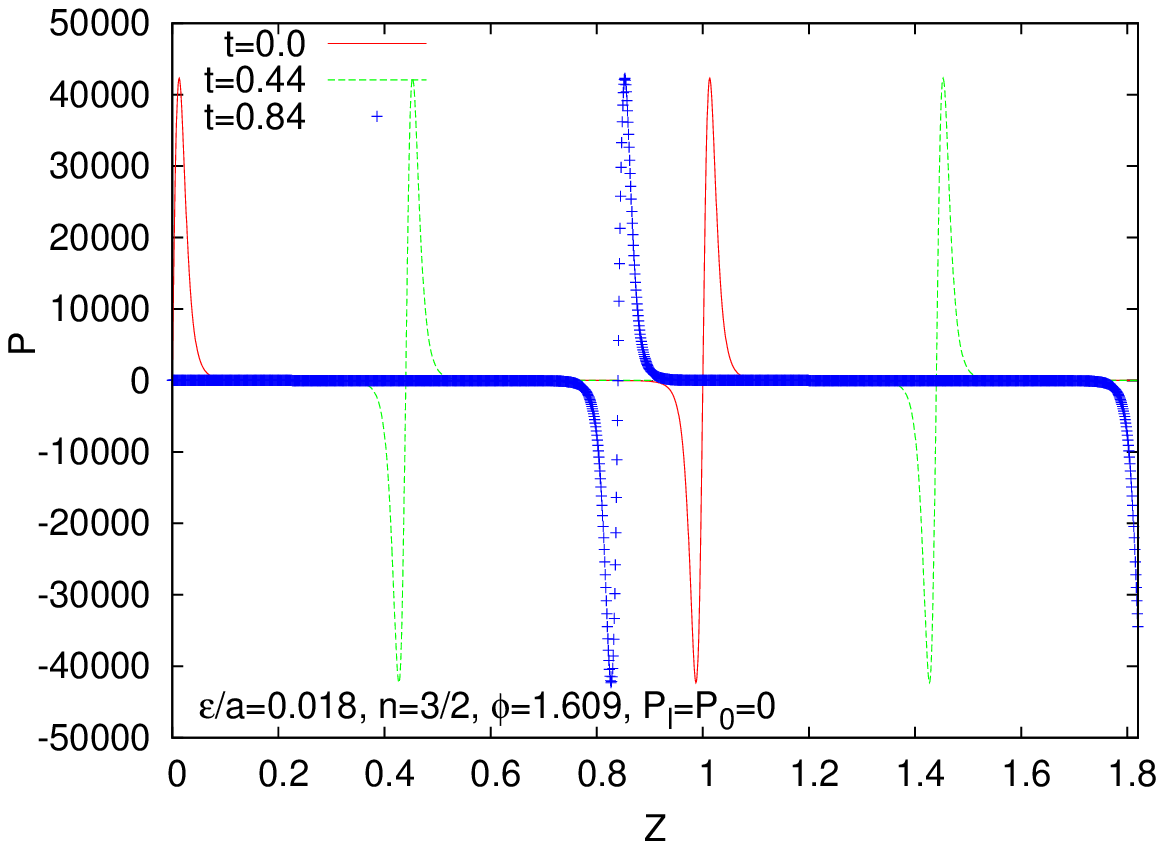}
\\$~~~~~~~~~~~~~~~~~~~~~~~~~~~(e)~~~~~~~~~~~~~~~~~~~~~~~~~~~~~~~~~~~~~~~~~~~~~~~~~~~~~~~~~~~~~~~~~~~~~(f)~~~~~~~~~~~~~~~$
\caption{Local pressure distribution in the esophagus in the case of a
  single wave. These findings match quite fairly with those of a
  Newtonian fluid reported earlier by Li and Brasseur \cite{Li1}. For
  a shear-thinning fluid (cf. (a-b):n=1/2) and a shear-thickening
  fluid (cf. (e-f): n=3/2) the overall behavior is found to be
  somewhat similar. Moreover, for both the cases of single wave and
  wave train propagation, while for a shear-thinning fluid (n=1/2),
  the magnitude of pressure is quite small (nearly 1.2-2.5$\%$ of a Newtonian fluid), for a shear-thickening fluid (n=3/2), it is quite large (nearly 260-300
times that for the Newtonian case).}
\label{paper6_pressurec6.2.1.1-6.2.3.4}
\end{figure}
In order to keep the fluid volume fixed within one wave period,
$\phi$ is adjusted when $\epsilon/a$ changes. When two waves are
present in the esophagus, Figs.
\ref{paper6_pressurec6.1.1.1-6.1.3.4}(c-d) give the pressure
variation of the fluid (considered Newtonian) at six specially
chosen locations during one wave period. The graphs indicate that
within one wave period, there are two peaks in the pressure
distribution within one wave period with a gradual pressure ramp in
between and transition occurs from large minimum peak to large
maximum peak at the point of maximum occlusion within the
contraction zone. It is further seen that to the right of this point
the wall of esophagus moves radially inward ($\partial H/\partial t
<0$) presumably owing to the contraction of the circular muscle. As
a result of this, a large pressure gradient is created there. To the
left of the point of minimum radius the wall moves radially
outwards. It causes a corresponding drop in pressure there.
Therefore local instantaneous motion occurs to the left of the point
of maximum occlusion and also to the right at the remaining portion
of the region. The net averaged flow over one wave period takes
place towards the wave. We find that the results obtained on the
basis of the present analysis and the form of $\bar{Q}(Z,t)$ given
by (\ref{paper6_volume_flow_assumed_form}) match with those reported
in \cite{Jaffrin2}. A comparison of variation of pressure between
Newtonian and rheological (non-Newtonian) fluids suggest that
pressure is highly sensitive to the rheological fluid index `n'.
Although the nature of pressure change along the tube length is
almost similar, it is noted from Figs.
\ref{paper6_pressurec6.1.1.1-6.1.3.4}(a-b) that the amount of change
is very small for a shear-thinning fluid with n=1/2. The change has
been observed clearly throughout the tube. For a shear-thickening
liquid with n=3/2, Figs. \ref{paper6_pressurec6.1.1.1-6.1.3.4}(e-f)
show that the magnitude of pressure is very large compared to that
for a Newtonian fluid.
\begin{figure}
\includegraphics[width=3.5in,height=2.1in]{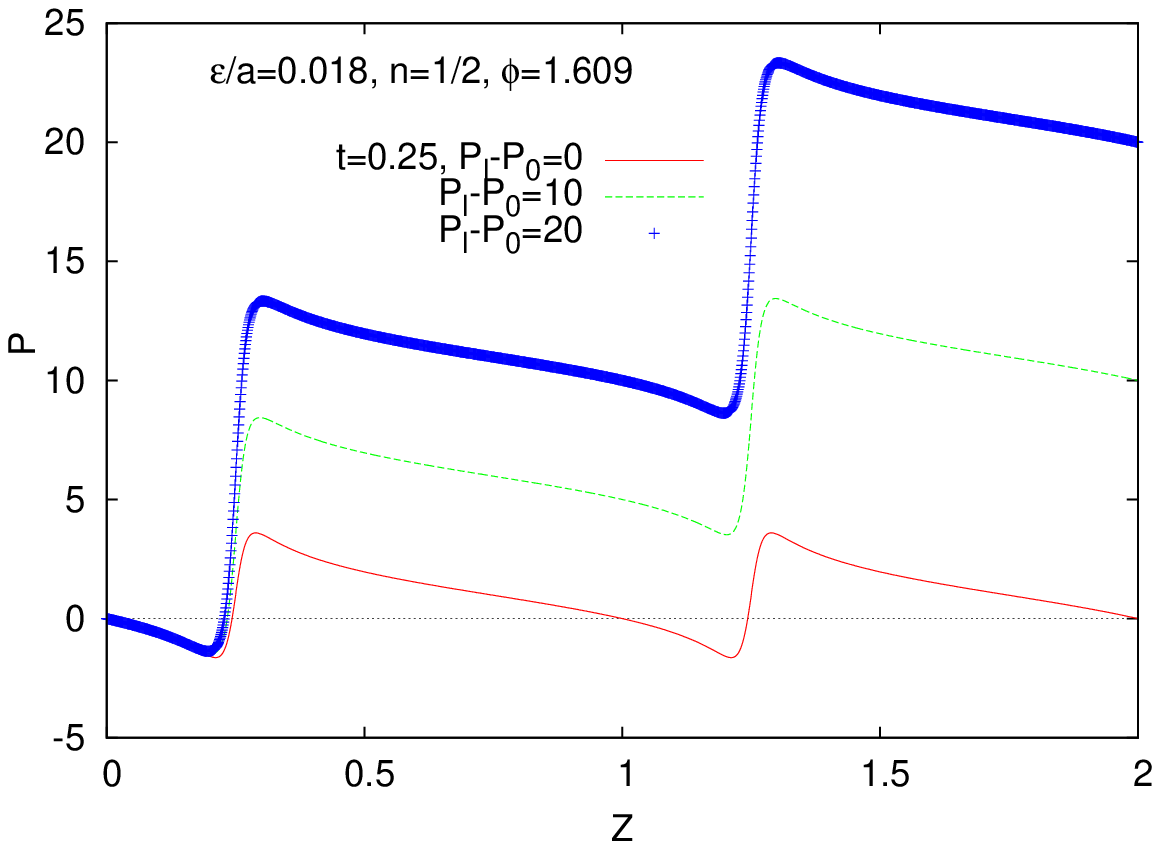}\includegraphics[width=3.5in,height=2.1in]{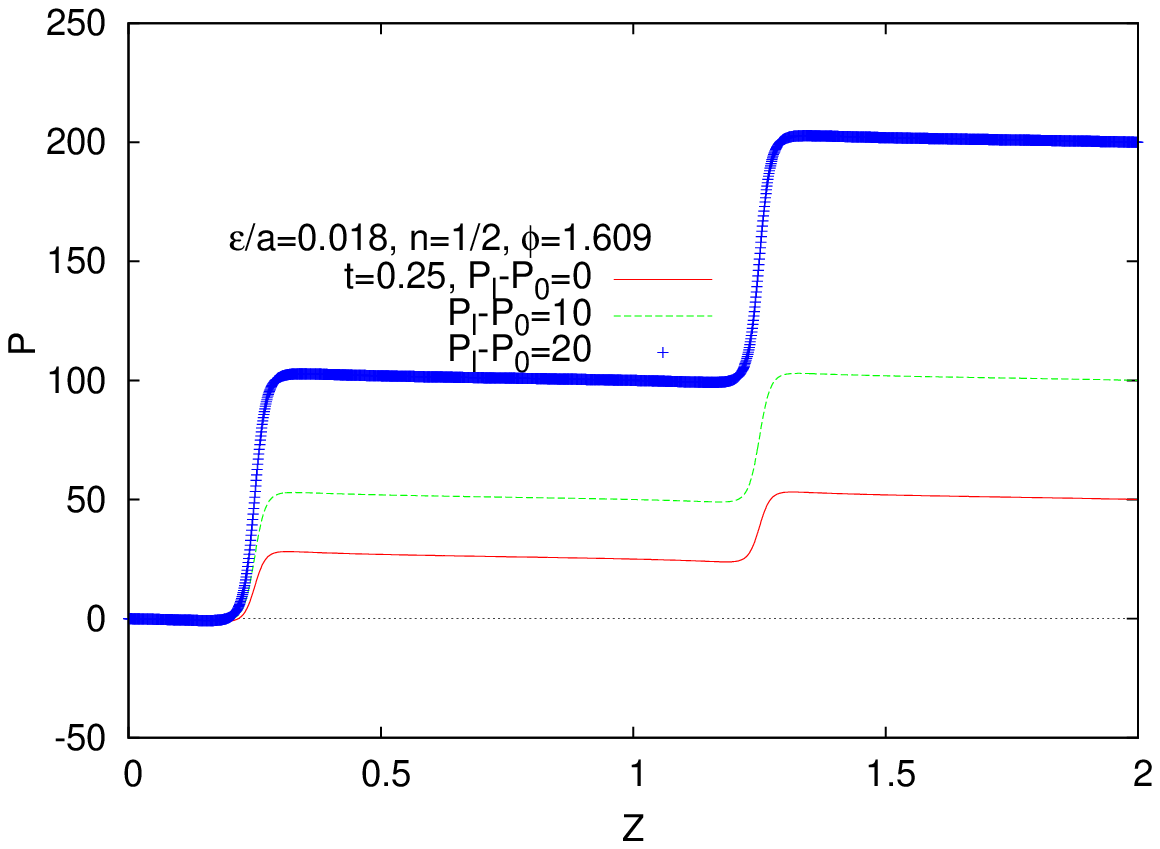}
\\$~~~~~~~~~~~~~~~~~~~~~~~~~~~(a)~~~~~~~~~~~~~~~~~~~~~~~~~~~~~~~~~~~~~~~~~~~~~~~~~~~~~~~~~~~~~~~~~~~~~(b)~~~~~~~~~~~~~~~$\\
\includegraphics[width=3.5in,height=2.1in]{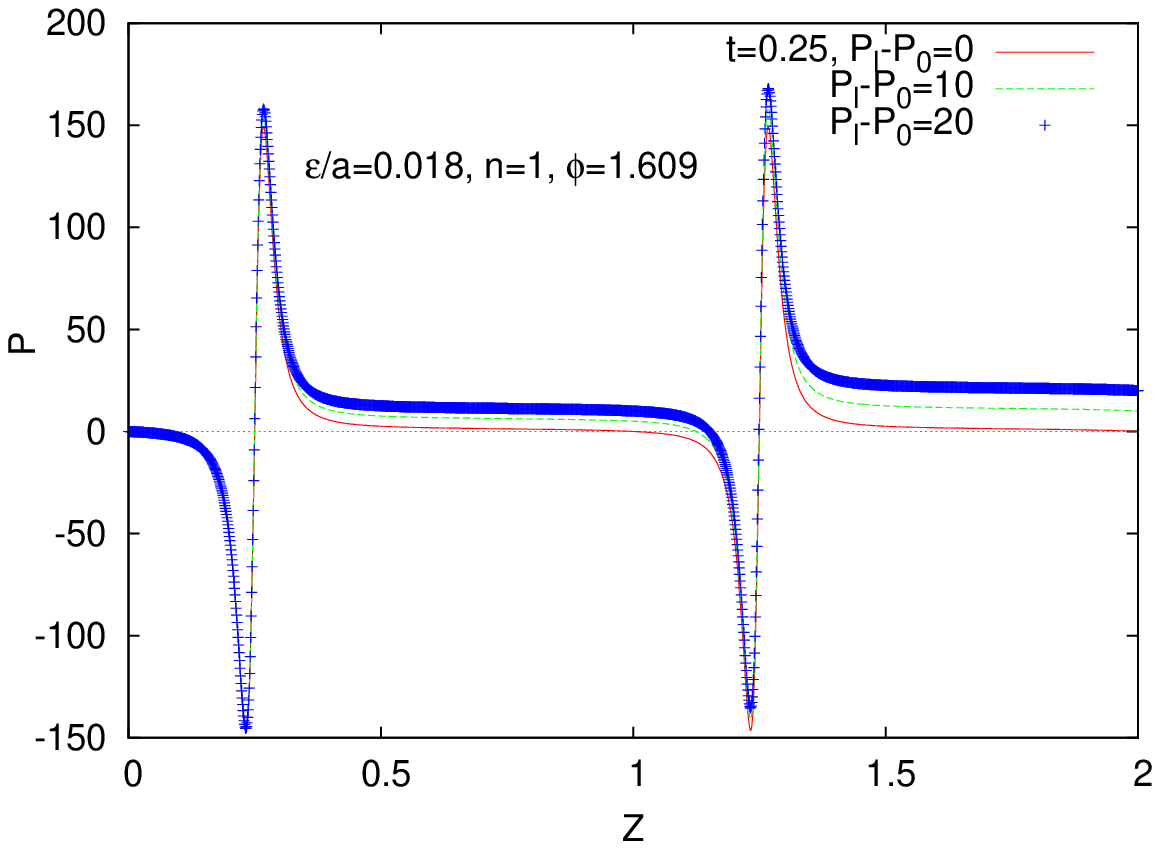}\includegraphics[width=3.5in,height=2.1in]{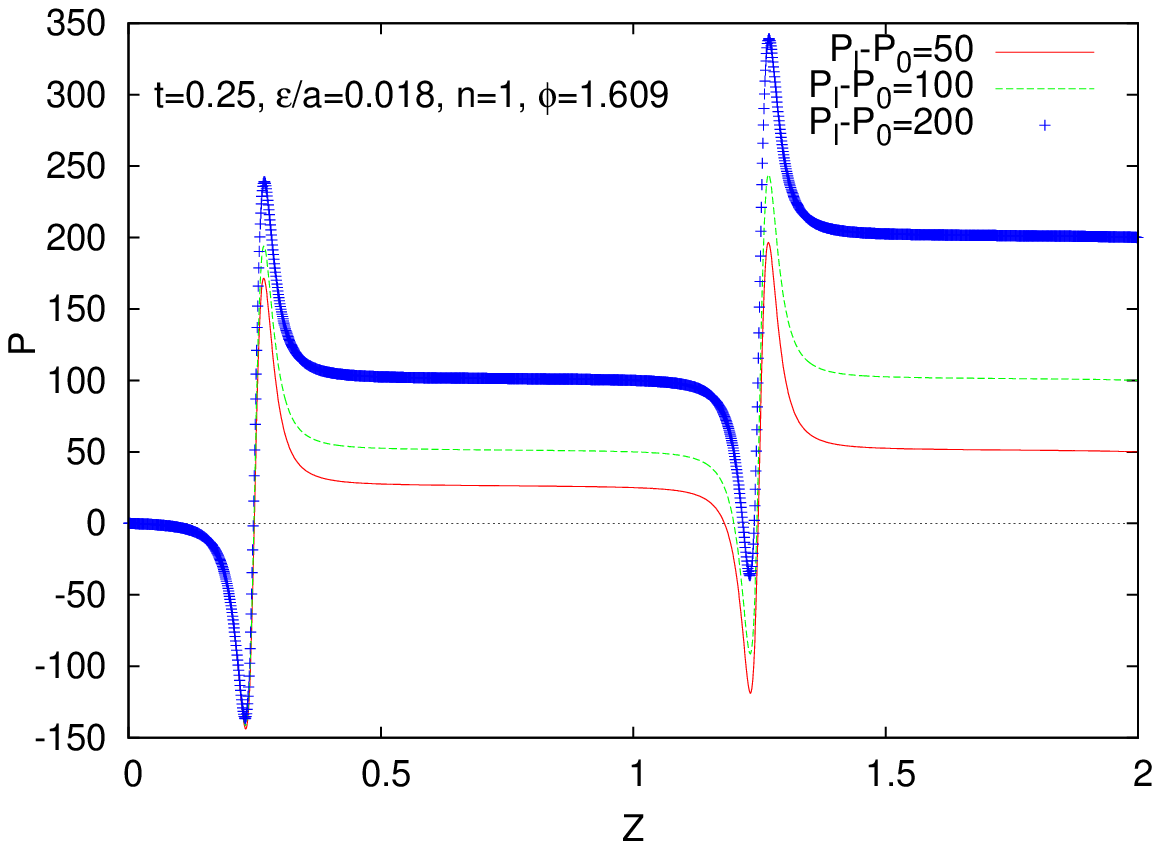}
\\$~~~~~~~~~~~~~~~~~~~~~~~~~~~(c)~~~~~~~~~~~~~~~~~~~~~~~~~~~~~~~~~~~~~~~~~~~~~~~~~~~~~~~~~~~~~~~~~~~~~(d)~~~~~~~~~~~~~~~$\\
\includegraphics[width=3.4in,height=2.1in]{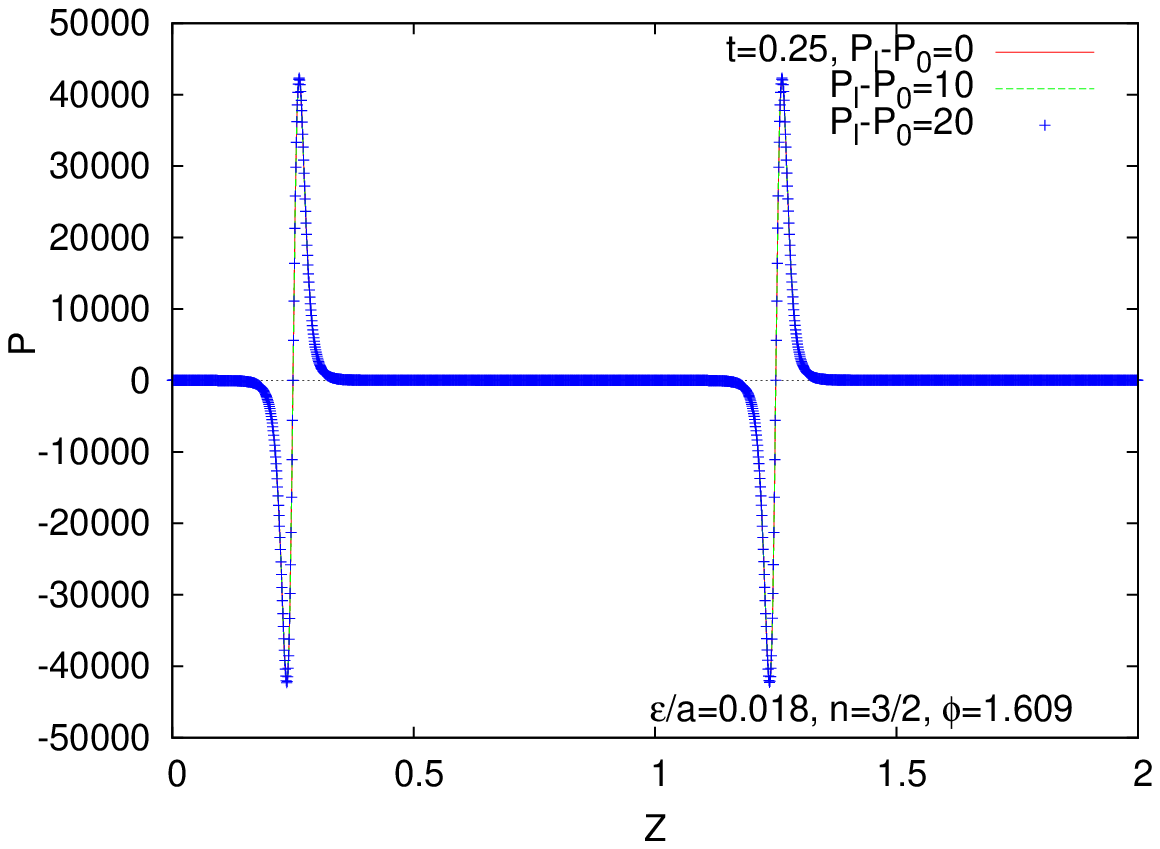}\includegraphics[width=3.5in,height=2.1in]{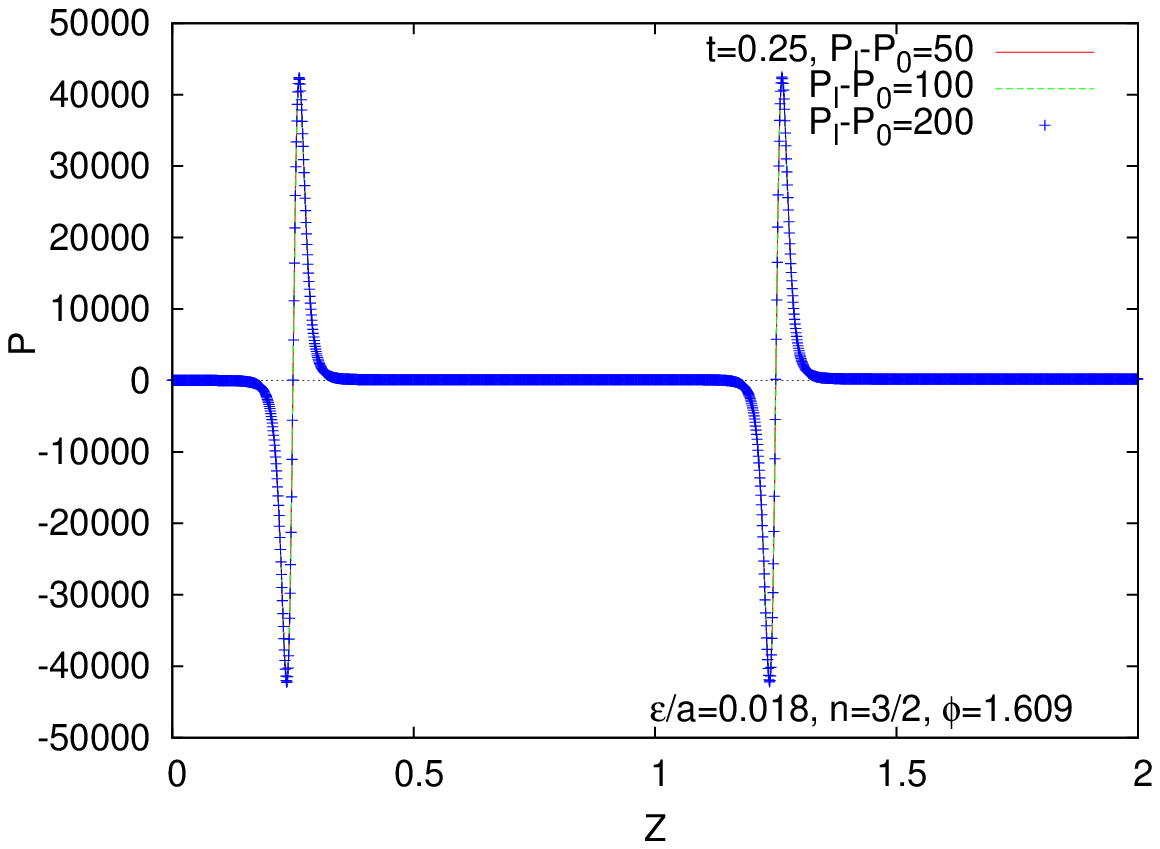}
\\$~~~~~~~~~~~~~~~~~~~~~~~~~~~(e)~~~~~~~~~~~~~~~~~~~~~~~~~~~~~~~~~~~~~~~~~~~~~~~~~~~~~~~~~~~~~~~~~~~~~(f)~~~~~~~~~~~~~~~$\\
\caption{Local pressure distribution in the case of a wave train
  propagation during movement of the food bolus. Pressure rise at the
  lower end of the esophagus (LEE) leads to an enhancement of pressure
  throughout the length of the esophagus.}
\label{paper6_pressurec6.1.4.1-6.1.9.1}
\end{figure}

In order to discuss the significant differences in reflux and and
pumping phenomena between single and wave train peristaltic
transport of rheological fluids, it is worthwhile to compare both
the spatial and temporal pressure variations. Figs.
\ref{paper6_pressurec6.2.1.1-6.2.3.4} present the comparison, where
the spatial variations in pressure are given at fixed times for
single bolus transport and train wave transport with a non-integral
number of waves in the esophagus. For the purpose of comparison of
the results of the present study for the rheological fluid
((non-Newtonian) with those for the Newtonian fluid, the results
obtained in \cite{Li1} are reproduced on the basis of our present
study for the Newtonian case in Figs.
\ref{paper6_pressurec6.2.1.1-6.2.3.4}(c-d). An extended adverse
pressure gradient for a Newtonian fluid (cf. Fig.
\ref{paper6_pressurec6.2.1.1-6.2.3.4}(c)) is found to be created by
the peristaltic wave from the inlet of the esophagus to its tail and
the outlet of the esophagus to the head of the peristaltic wave.
Thereby the motion is opposite to that of the peristaltic wave in
the said region. Fig. \ref{paper6_pressurec6.2.1.1-6.2.3.4}(d) shows
that in the wave train case, these extended regions are absent and
during its transit to the outlet of the esophagus a single
peristaltic wave is followed by an ever-increasing region of
backward motion. Some portion of the backward motion remains at the
outlet until the peristaltic wave head reaches the outlet. In the
contrary, when the bolus passes through the outlet of the esophagus,
it carries the fluid along with it that leads the net transport in
the direction of the wave.
\begin{figure}
\includegraphics[width=3.5in,height=2.1in]{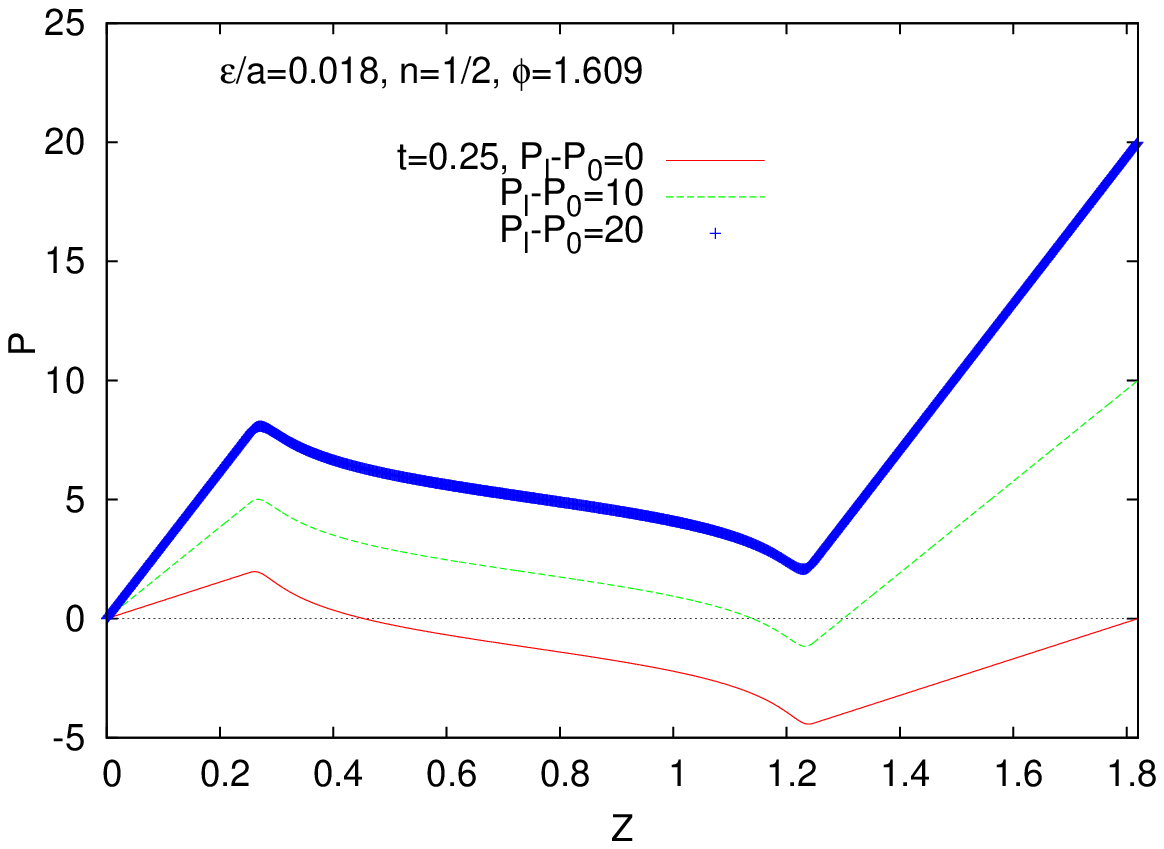}\includegraphics[width=3.5in,height=2.1in]{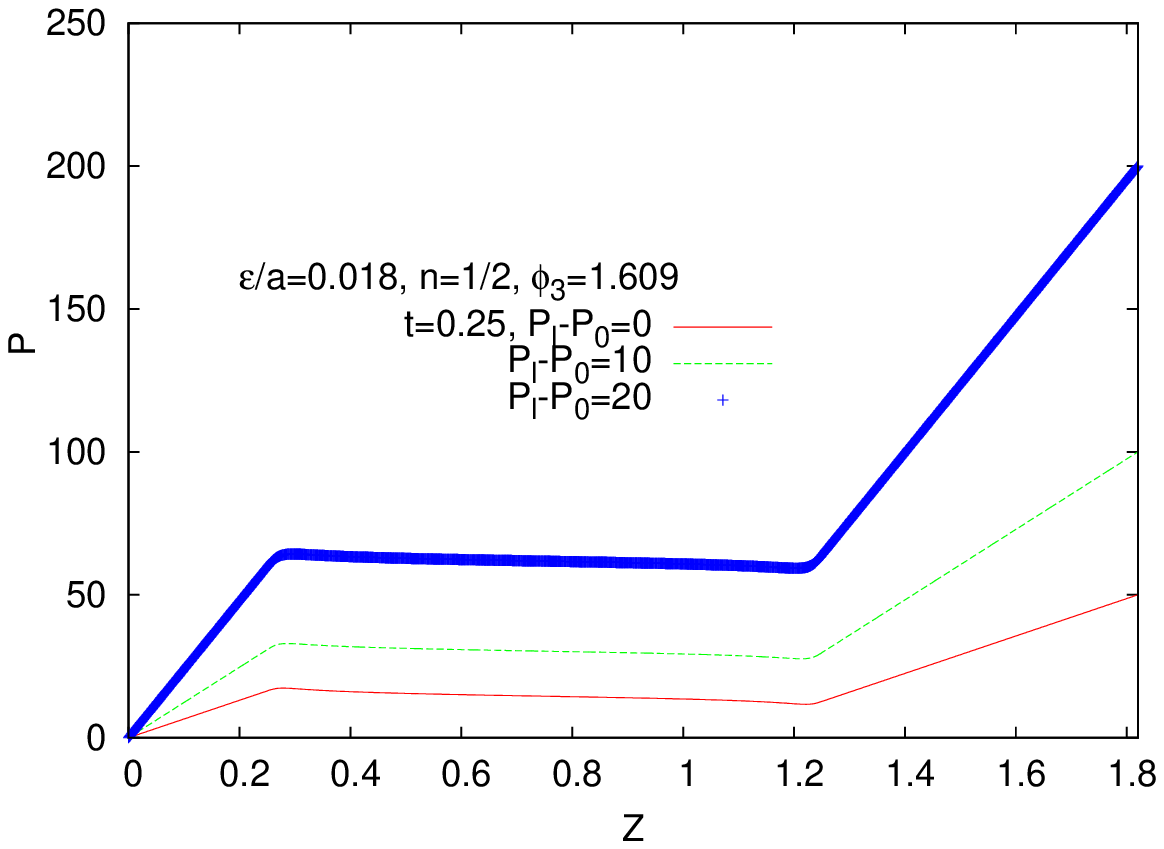}
\\$~~~~~~~~~~~~~~~~~~~~~~~~~~~(a)~~~~~~~~~~~~~~~~~~~~~~~~~~~~~~~~~~~~~~~~~~~~~~~~~~~~~~~~~~~~~~~~~~~~~(b)~~~~~~~~~~~~~~~$\\
\includegraphics[width=3.5in,height=2.1in]{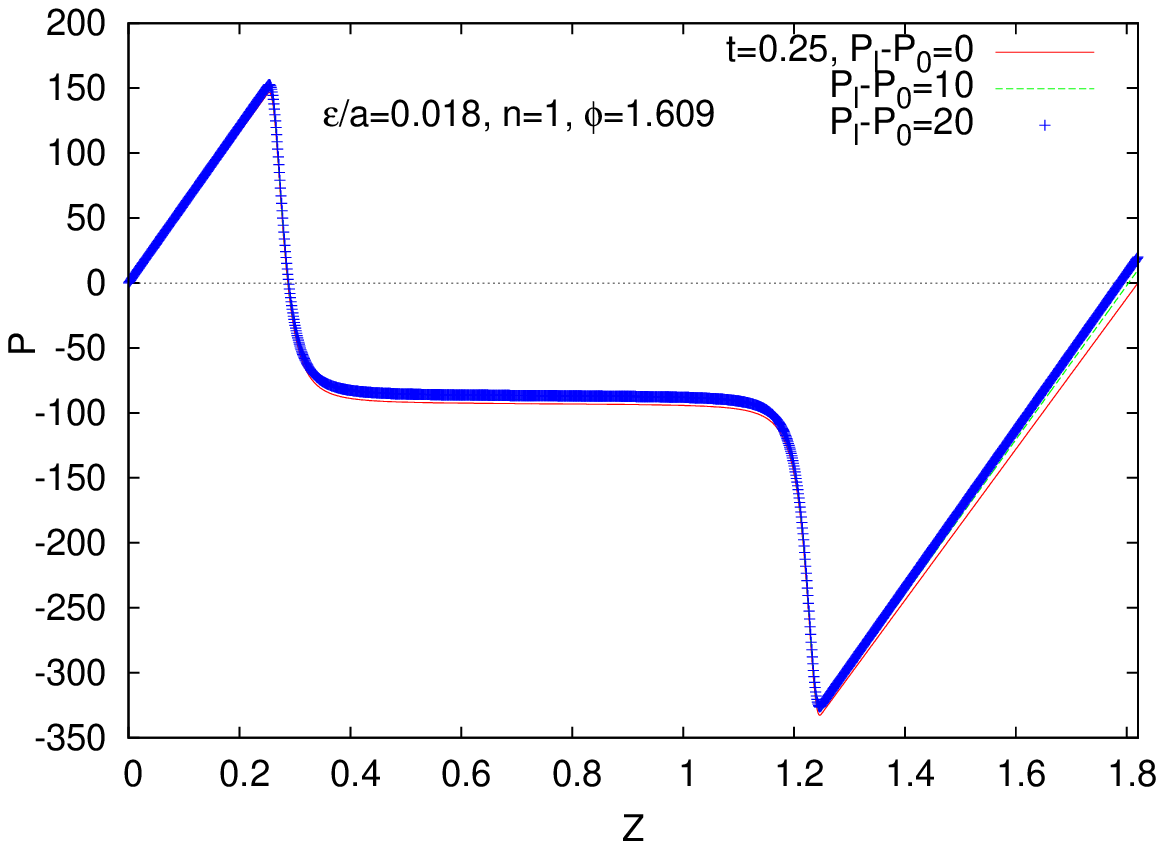}\includegraphics[width=3.5in,height=2.1in]{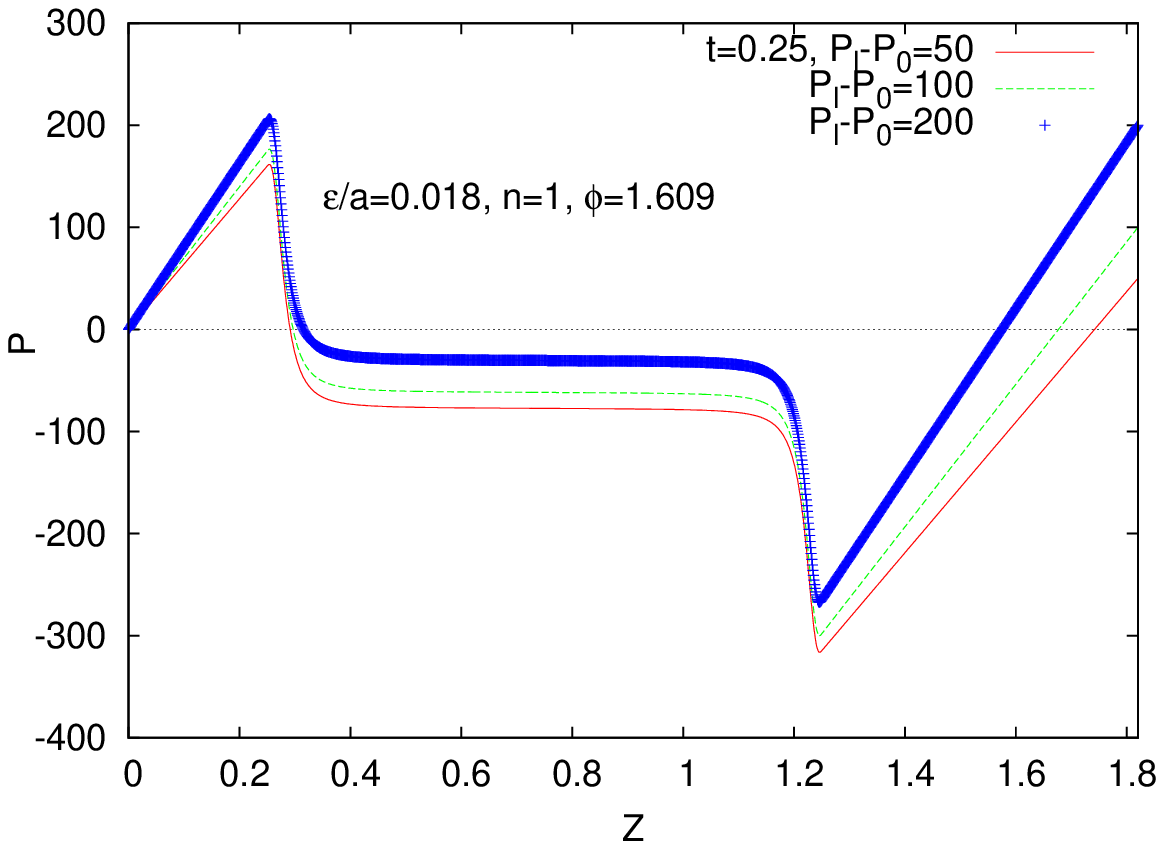}
\\$~~~~~~~~~~~~~~~~~~~~~~~~~~~(c)~~~~~~~~~~~~~~~~~~~~~~~~~~~~~~~~~~~~~~~~~~~~~~~~~~~~~~~~~~~~~~~~~~~~~(d)~~~~~~~~~~~~~~~$\\
\includegraphics[width=3.4in,height=2.1in]{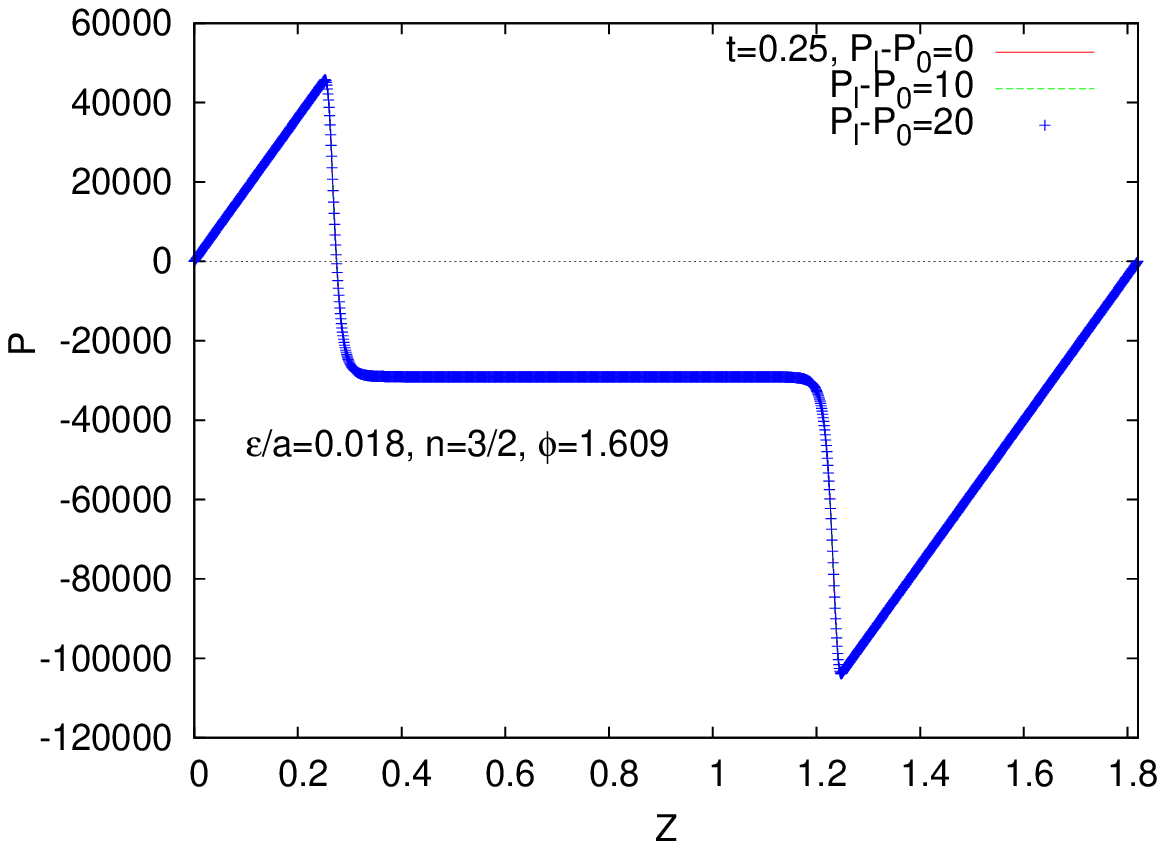}\includegraphics[width=3.5in,height=2.1in]{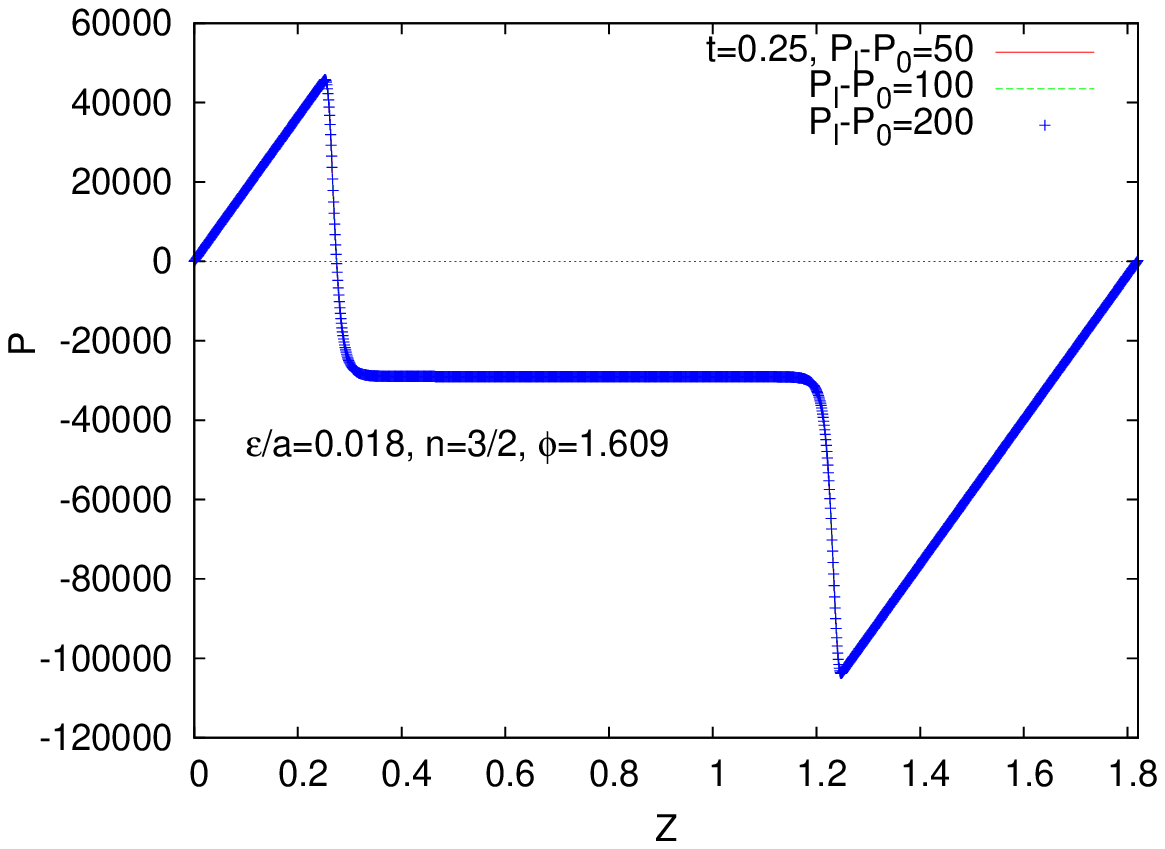}
\\$~~~~~~~~~~~~~~~~~~~~~~~~~~~(e)~~~~~~~~~~~~~~~~~~~~~~~~~~~~~~~~~~~~~~~~~~~~~~~~~~~~~~~~~~~~~~~~~~~~~(f)~~~~~~~~~~~~~~~$\\
\caption{Local pressure distribution in the case of a single wave
  during food bolus movement. As pressure at the lower end of
  esophagus (LEE) rises, region of adverse pressure gradient is
  extended rapidly for a shear thinning fluid, while for a Newtonian
  fluid it is significant only after the pressure at LEE ($P_l$)
  exceeds a critical value.}
\label{paper6_pressurec6.2.4.1-6.2.9.1}
\end{figure}

It is noticed that unlike single wave propagation, wave train has as
many pairs of peaks of pressure in both Newtonian and rheological
fluids as the number of waves present in the duct in a given time
interval. Moreover, in case of train wave propagation, the pressure
transit at once from minimum to maximum when the wave head is
immediately succeeded by the tail of the leading wave for all types
of fluids examined here.

A negative pressure difference drives a positive flow, whereas a
positive pressure difference creates the resistance of the flow.
When it attains a certain critical value (that depends on the wave
amplitude), the power-law index and other related conditions, there
is a possibility that the flow would be completely restrained. If it
exceeds that critical value, the flow will take place in the
backward direction. This causes emesis (in clinical terms), which is
commonly known as vomiting that involves forceful expulsion of the
contents of the stomach through the esophagus. Physiologically it
occurs due to gastritis, or poisoning, or due to high intracranial
pressure or over exposure to conizing radiation. This may be also
happen to patients suffering from brain tumor. The backward flow of
undigested   food from the stomach to the mouth is, however, called
medically as regurgitation. Figs.
\ref{paper6_pressurec6.1.4.1-6.1.9.1} and Figs.
\ref{paper6_pressurec6.2.4.1-6.2.9.1} exhibit local pressure
distribution when the pressure at the lower end is greater than that
at upper end of the esophagus for wave train transport and for
single bolus transport respectively. For a shear thinning fluid
Figs. \ref{paper6_pressurec6.1.4.1-6.1.9.1}(a-b), it is seen that
local pressure enhances significantly with the increase in
$\Delta$P.  For a Newtonian fluid (cf. Figs.
\ref{paper6_pressurec6.1.4.1-6.1.9.1}(c-d)), it is also increases
with the increase in $\Delta$P except at the transition region.
However, in the case of shear thickening fluid (n=3/2), Figs.
\ref{paper6_pressurec6.1.4.1-6.1.9.1}(e-f) indicate that value of
$\Delta$P considered here ( i.e. $0\leq\Delta P <200$) does not
significantly affect the pressure. In the case of a single bolus
transport, it is noted from Figs.
\ref{paper6_pressurec6.2.4.1-6.2.9.1} that as the pressure at the
lower end of esophagus (LEE)  increases, local pressure throughout
the region also increases for shear thinning fluid when n=1/2, where
as for Newtonian fluid this increase is significant when $P_1$
reaches a greater value.
\begin{figure}
\includegraphics[width=3.5in,height=2.0in]{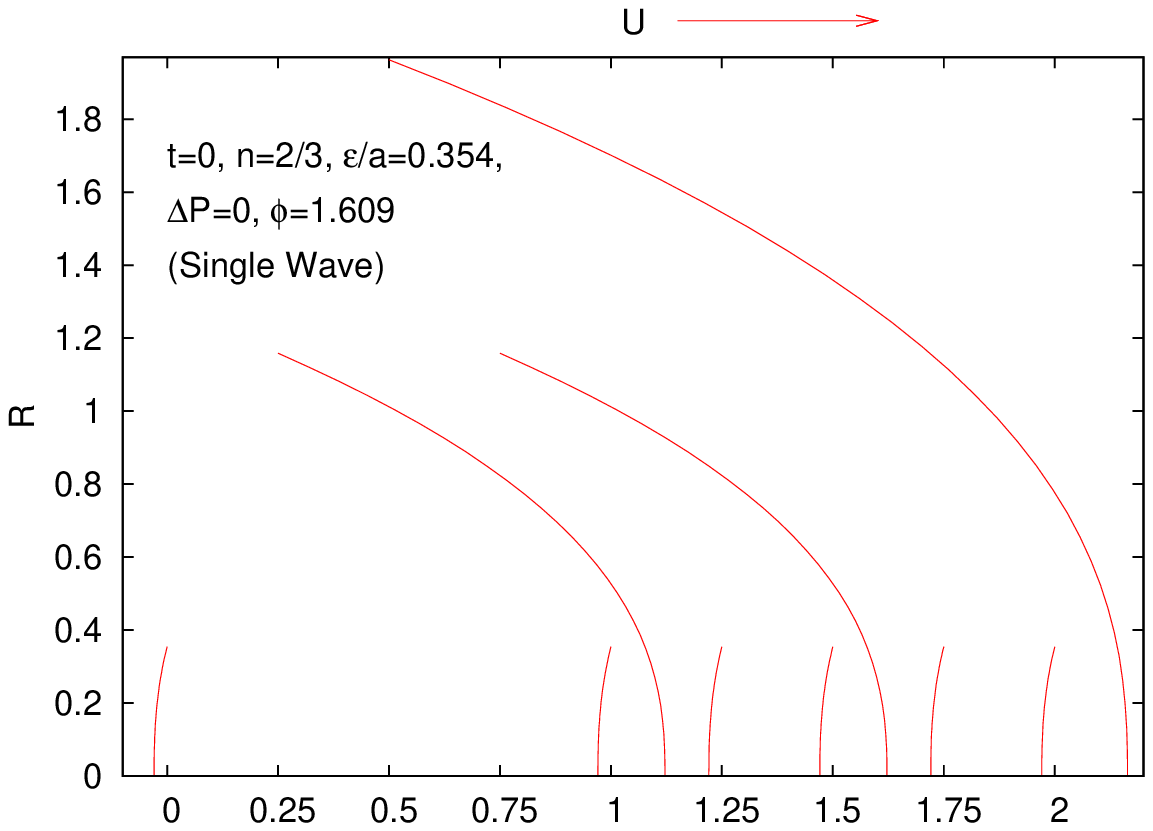}\includegraphics[width=3.5in,height=2.0in]{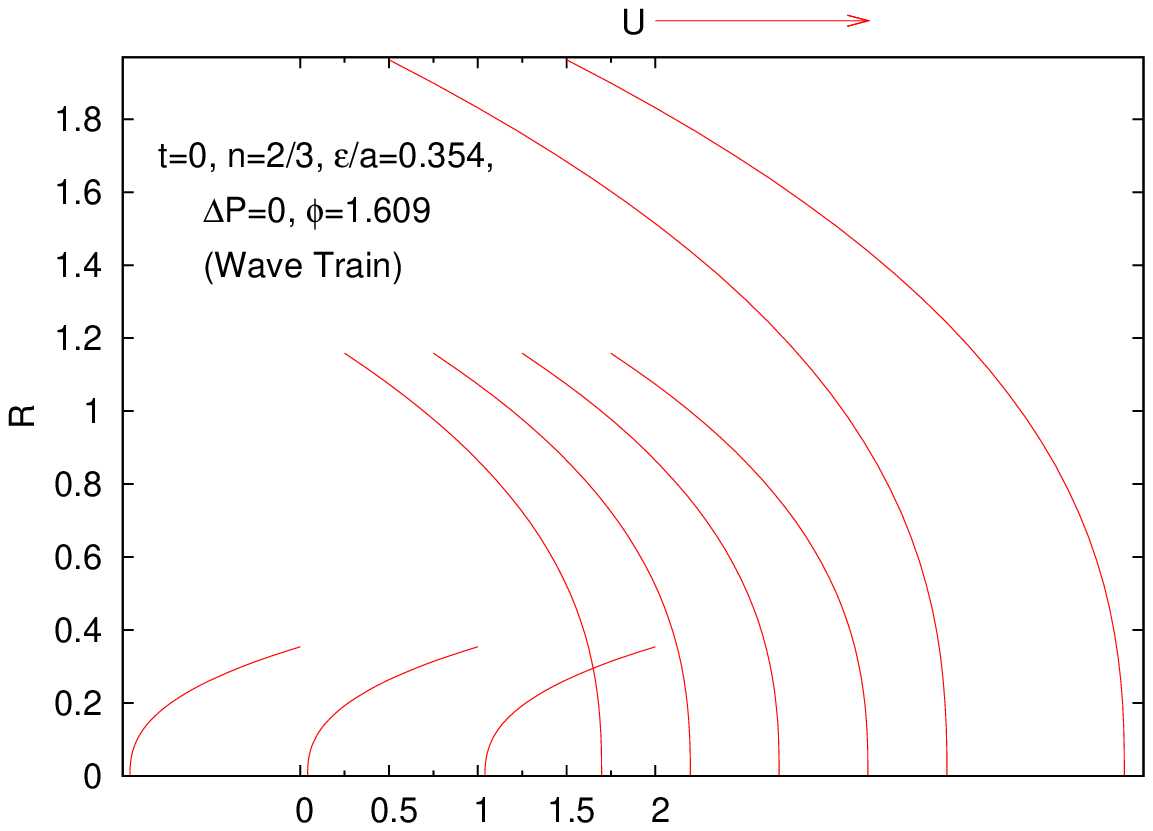}
\\$~~~~~~~~~~~~~~~~~(a)~~~~~~~~~~~~~~~~~~~~~~~~~~~~~~~~~~~~~~~~~~~~~~~~~~~~~~~~~~~~~~~~~~~(b)~~~~~~~~~$
\includegraphics[width=3.5in,height=2.0in]{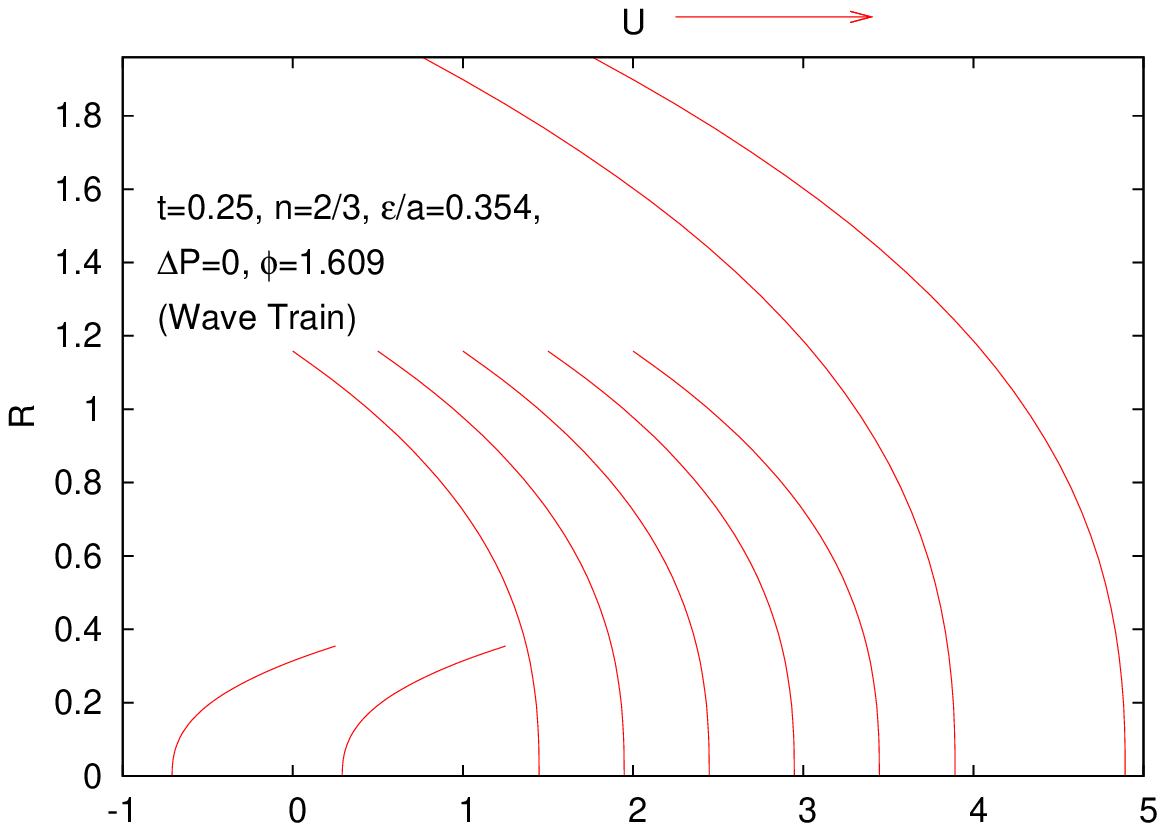}\includegraphics[width=3.5in,height=2.0in]{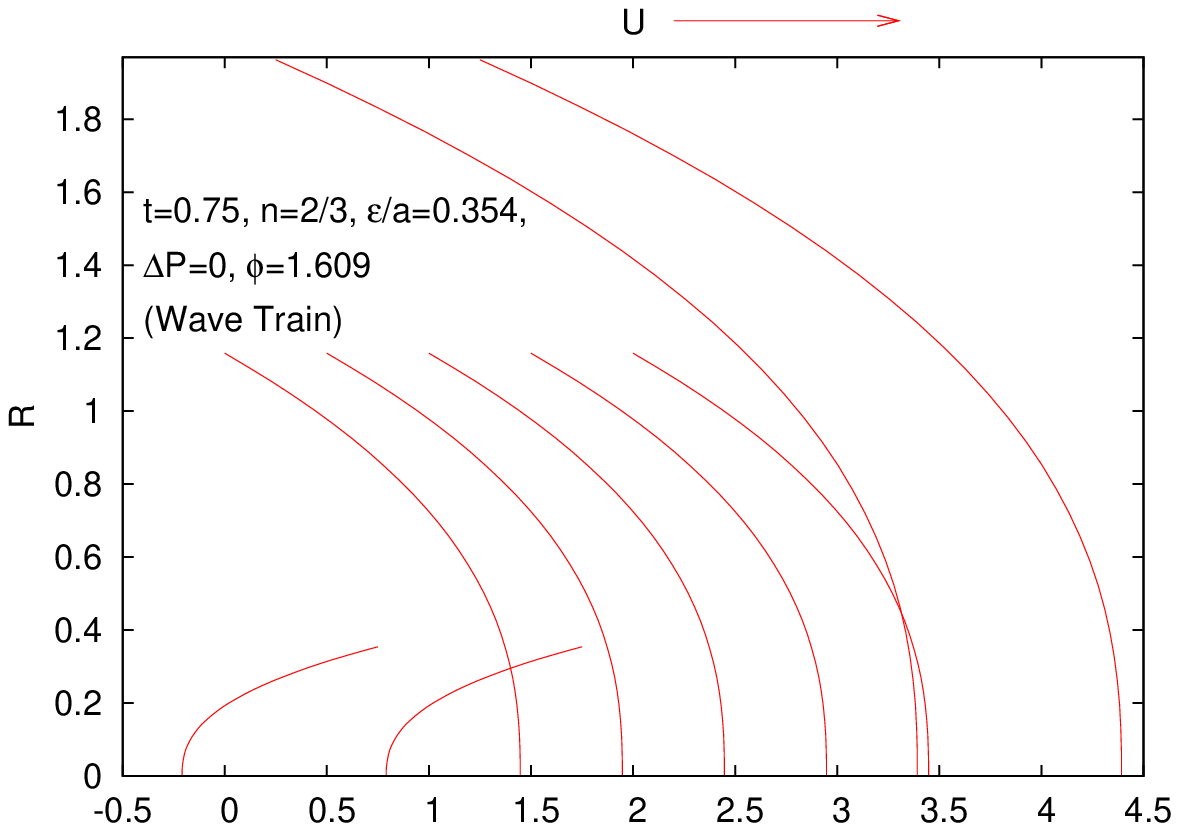}
\\$~~~~~~~~~~~~~~~~~(c)~~~~~~~~~~~~~~~~~~~~~~~~~~~~~~~~~~~~~~~~~~~~~~~~~~~~~~~~~~~~~~~~~~~(d)~~~~~~~~~$
\includegraphics[width=3.5in,height=2.0in]{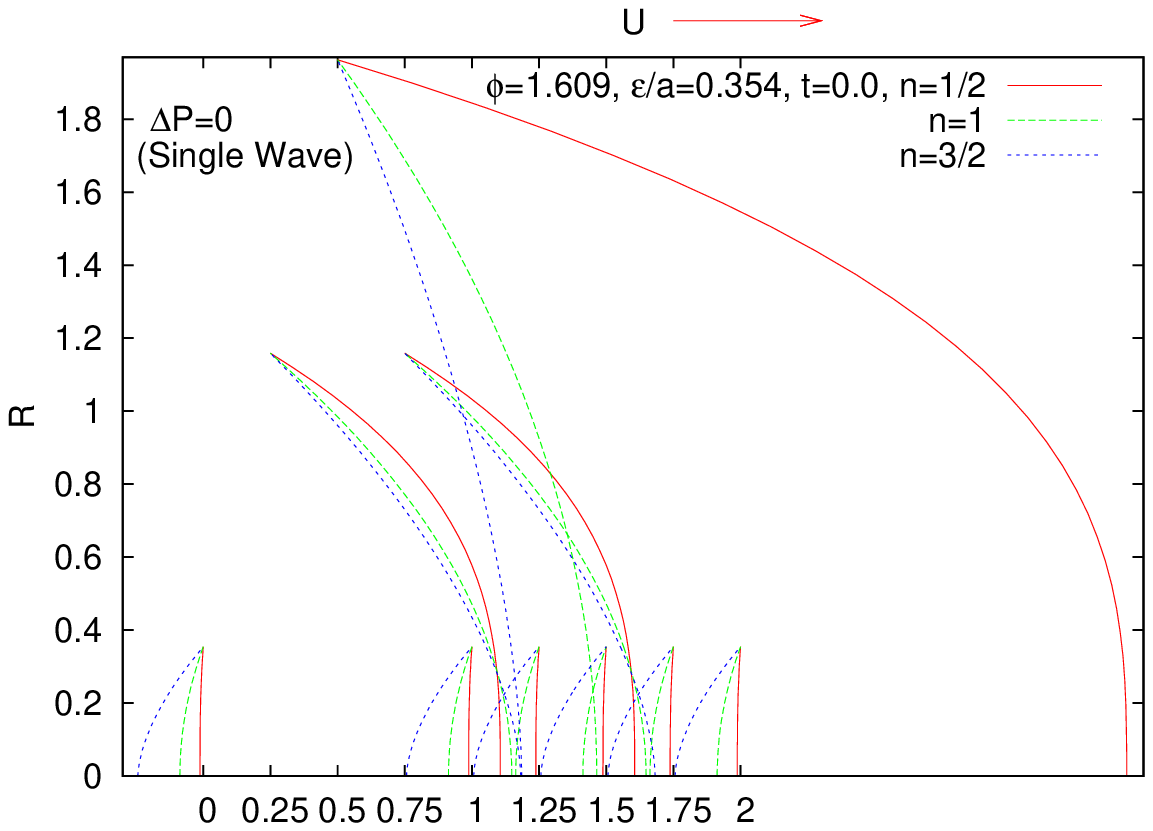}\includegraphics[width=3.5in,height=2.0in]{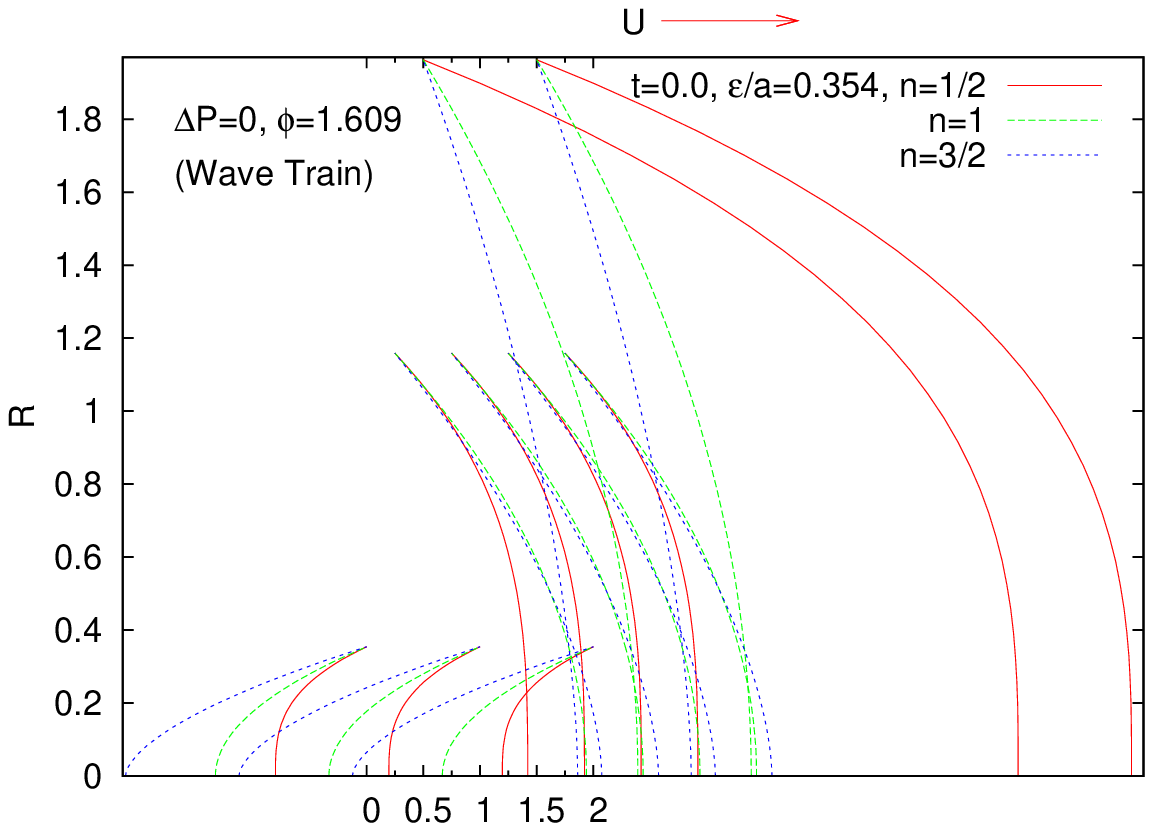}
\\$~~~~~~~~~~~~~~~~~(e)~~~~~~~~~~~~~~~~~~~~~~~~~~~~~~~~~~~~~~~~~~~~~~~~~~~~~~~~~~~~~~~~~~~(f)~~~~~~~~~$
\caption{Distribution of axial velocity at different instants of
time. Fig. (a) shows that in the region where the tube radius is minimum,
  the flow takes place in opposite direction but the magnitude of
  velocity is small, whereas in the remaining region the magnitude of
  the velocity is large and direction of velocity is the same as that
  of the propagating wave. Thus in the case of a single wave when the
  esophagus fails to maintain total occlusion, fluid transport takes
  place in the forward direction where the wave is active, while in
  other parts of the tube, the bolus moves slowly in the backward
  direction.}
\label{paper6_veloc6.1.1-6.2.5}
\end{figure}

\subsection{Velocity Distribution}
Since the velocity profiles, the pressure and the esophageal radius,
all change with time, it is pertinent to investigate the distribution
of velocity at different time intervals of a wave period. Moreover,
for a single wave transport, the limited region where peristaltic wave
is active deserve special attention. In the case of free pumping
($\Delta P=0$) for a single wave at t=0.0,
Fig. \ref{paper6_veloc6.1.1-6.2.5}(a) shows that in the region where
the tube radius is minimum, the flow takes place in opposite direction
but the magnitude of velocity is small, whereas in the remaining
region the magnitude of the velocity is large and direction of
velocity is the same as that of the propagating wave. Thus in the case
of a single wave when the esophagus fails to maintain total occlusion,
fluid transport takes place in the forward direction where the wave is
active, while in other parts of the tube, the bolus moves slowly in
the backward direction. As time progresses, although this trend is
maintained, the regions in which forward and backward flows occur,
change depending on the current position of the single wave. In the
contrary, for a wave train, the transport takes place with very high
velocity in both the forward and backward regions
(cf. Fig. \ref{paper6_veloc6.1.1-6.2.5}(b)). It may be noted that
backward flow occurs mainly between the junction of the wave lengths,
although occurrence of forward and backward flow regions is similar to
the single wave case. However, some difference in the regions of
forward and backward velocity profiles of forward and backward motions
is observed with the passage of time (cf. Figs
\ref{paper6_veloc6.1.1-6.2.5}(b-d)). Figs. \ref{paper6_veloc6.1.1-6.2.5}(e-f) indicate that as the
fluid index number `n' increases, backward flow is enhanced, while the
forward flow reduces significantly, whether it is a case of single
wave propagation or that of a wave train propagation.

\begin{figure}
\begin{center}
\includegraphics[width=3.5in,height=2.0in]{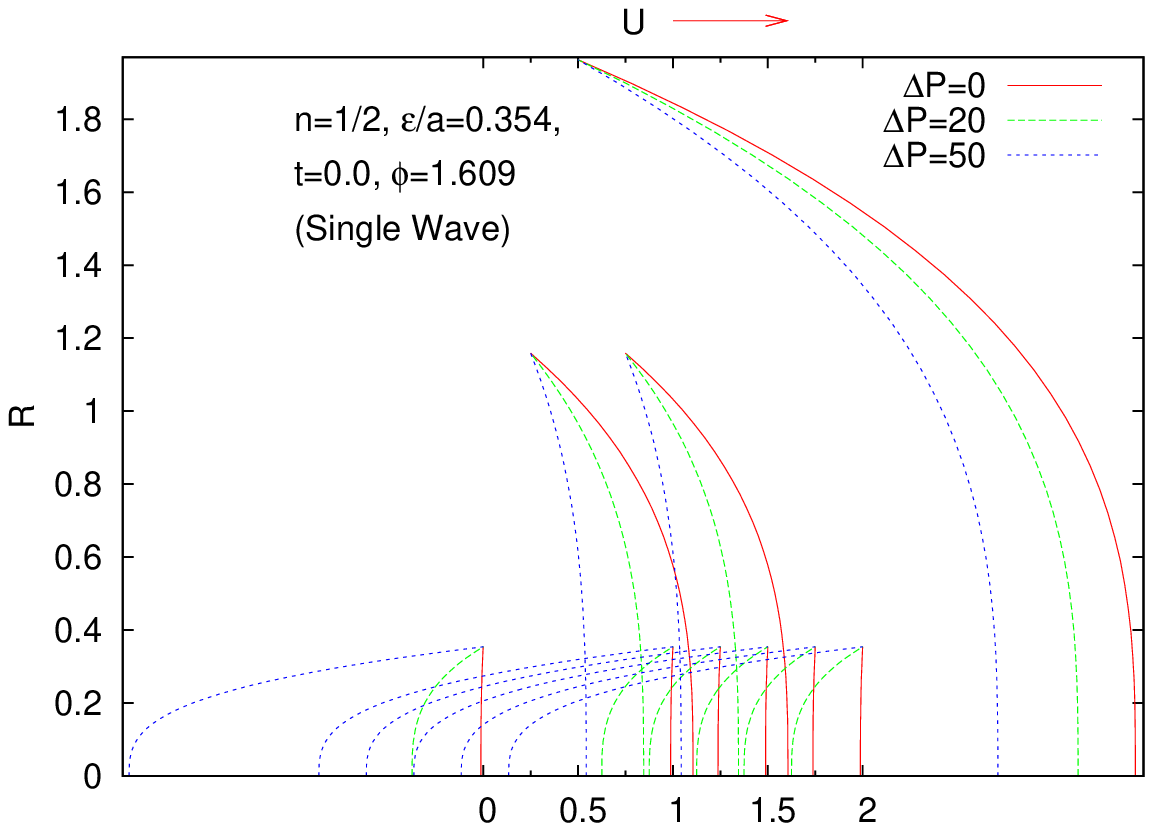}\includegraphics[width=3.5in,height=2.0in]{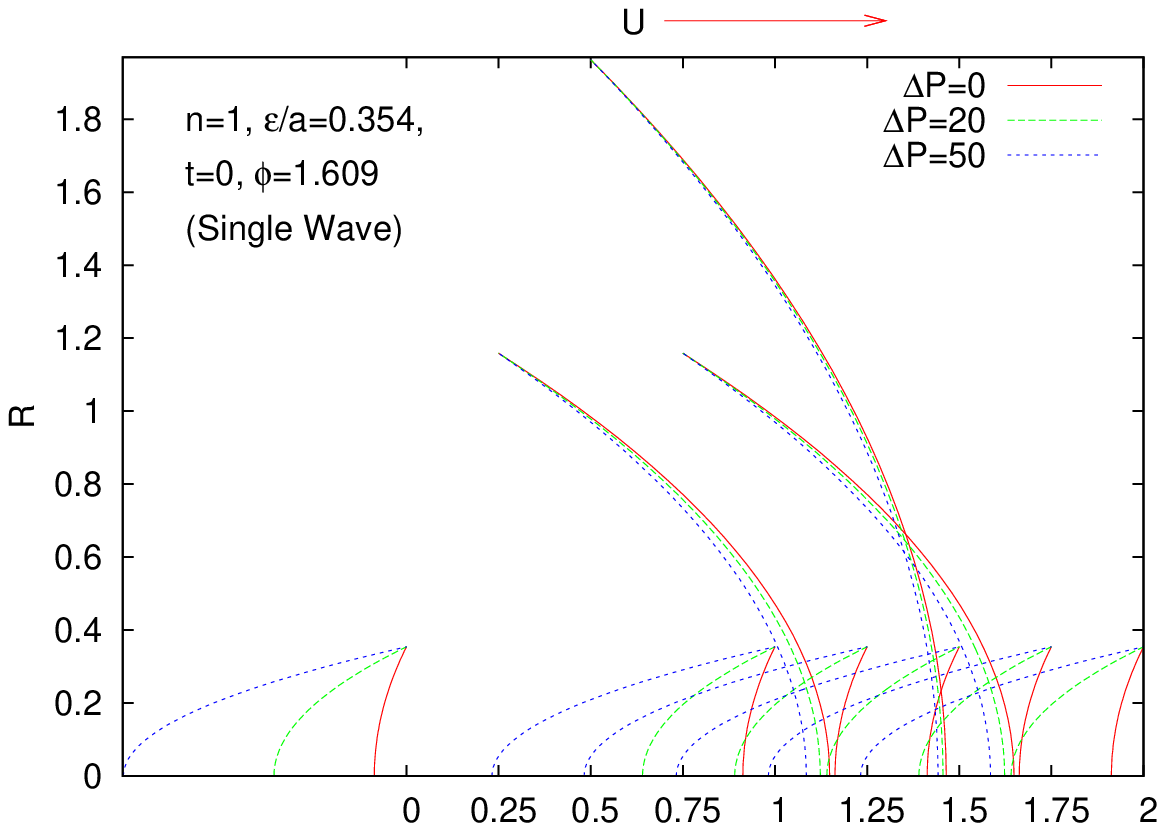}
\\$~~~~~~~~~~~~~~~~~(a)~~~~~~~~~~~~~~~~~~~~~~~~~~~~~~~~~~~~~~~~~~~~~~~~~~~~~~~~~~~~~~~~~~~(b)~~~~~~~~~$
\includegraphics[width=3.5in,height=2.0in]{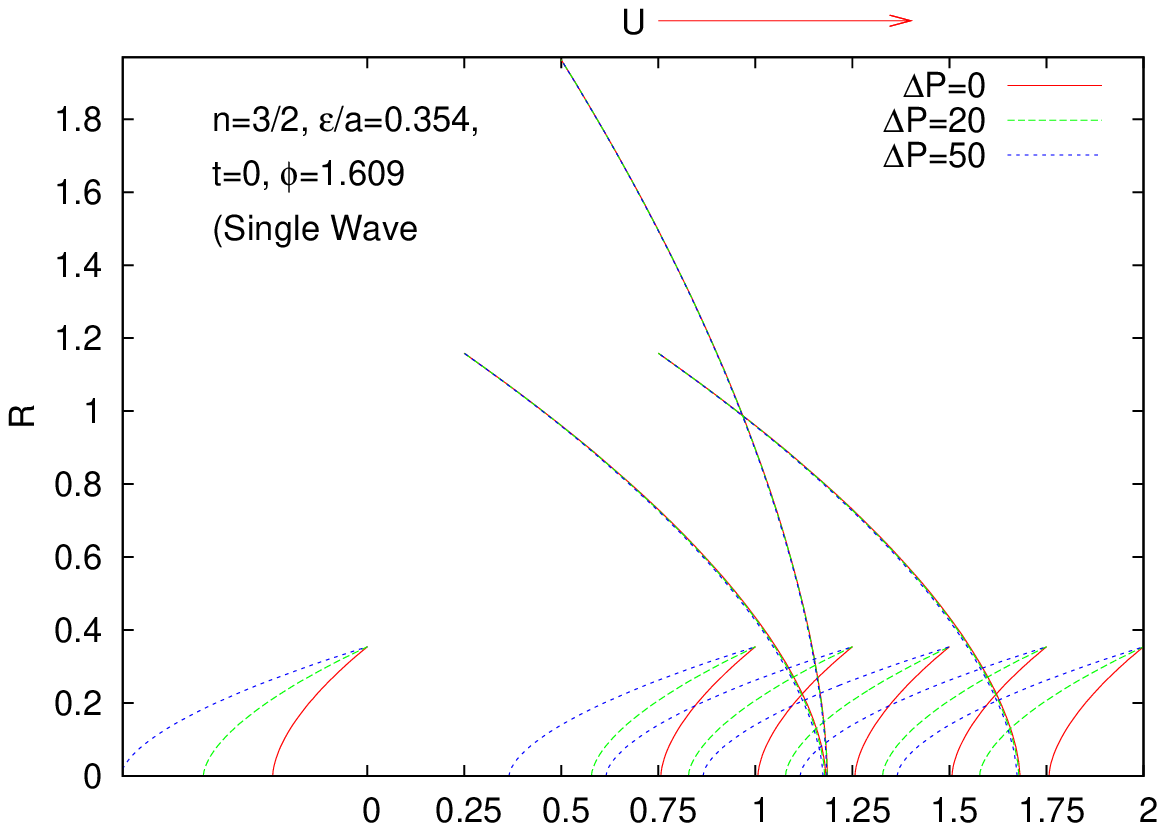}
\\(c)
\caption{In the case of a large pressure gradient at the two ends of
  the tube, where the pressure at the lower end is higher, the reflux
  region is prominent in the case of a single bolus transport. This is
  in the contrast to the case when the pressure gradient is zero.}
\label{paper6_veloc6.1.9-6.1.15}
\end{center}
\end{figure}

\begin{figure}
\begin{center}
\includegraphics[width=3.5in,height=2.0in]{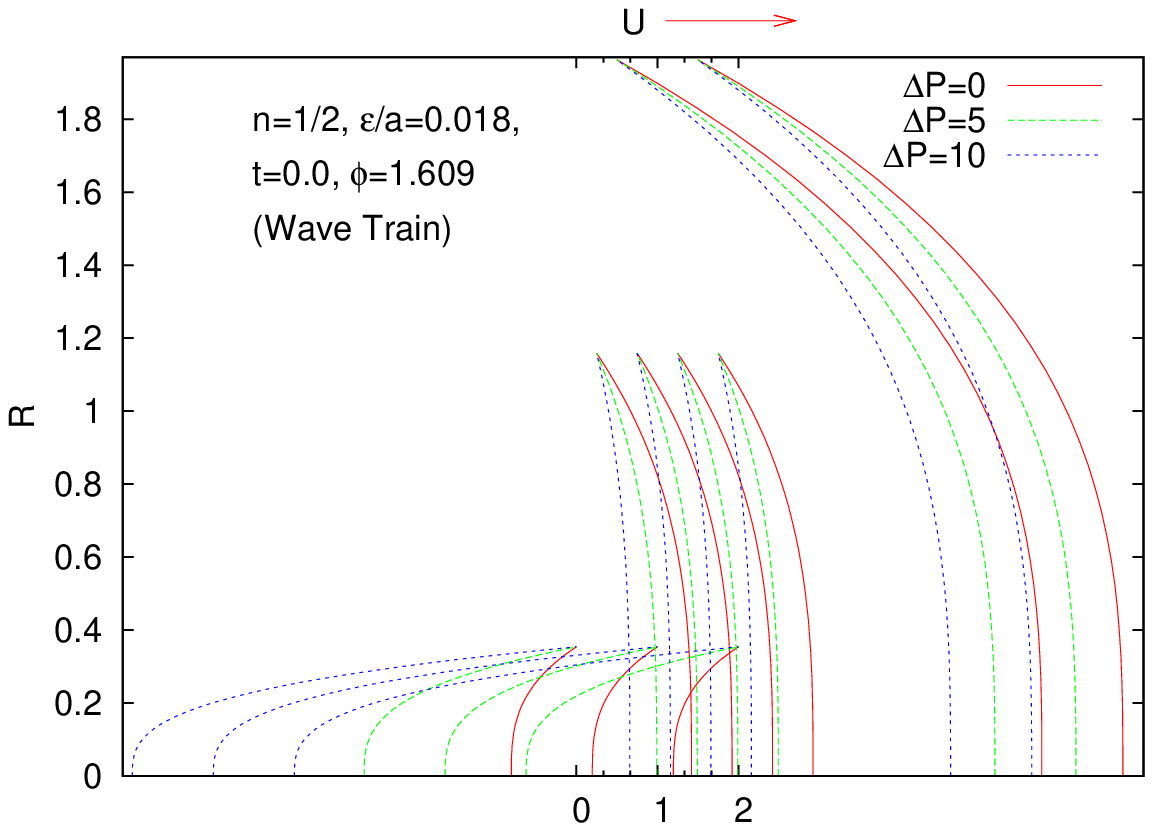}\includegraphics[width=3.5in,height=2.0in]{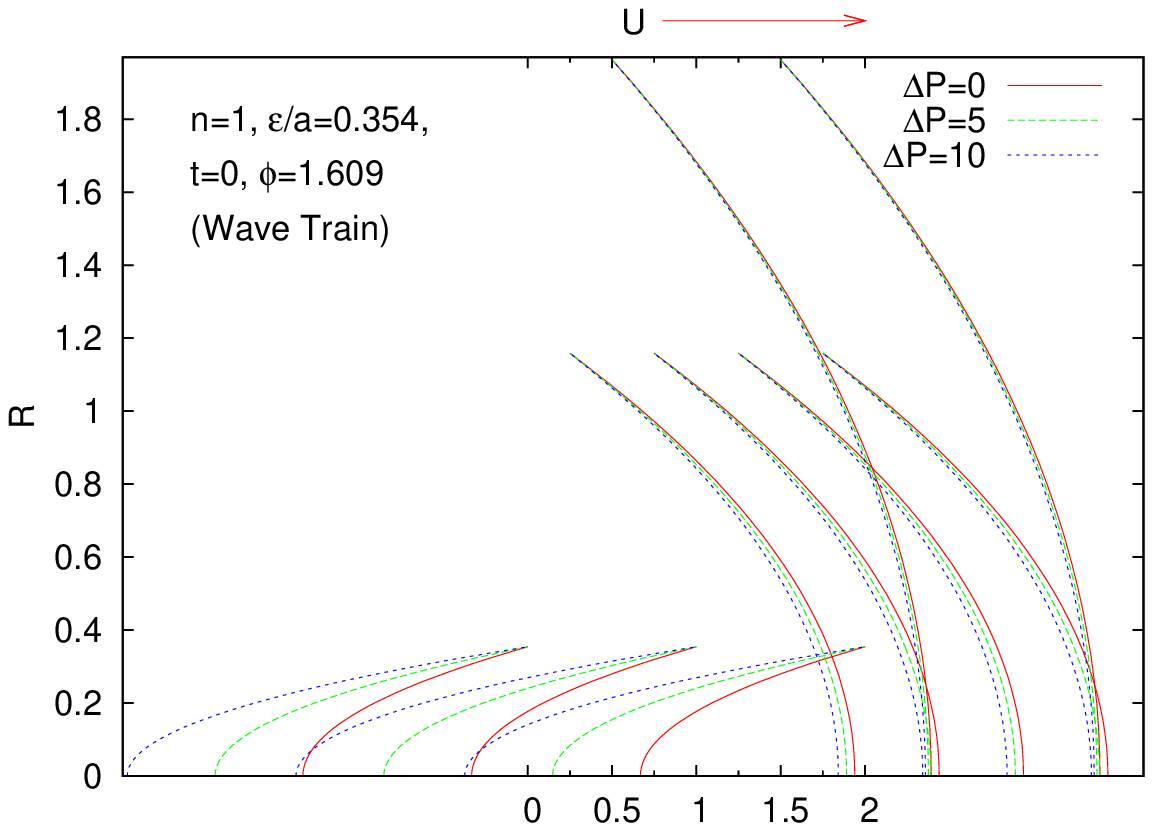}
\\$~~~~~~~~~~~~~~~~~(a)~~~~~~~~~~~~~~~~~~~~~~~~~~~~~~~~~~~~~~~~~~~~~~~~~~~~~~~~~~~~~~~~~~~(b)~~~~~~~~~$
\includegraphics[width=3.5in,height=2.0in]{veloc6.1.15.eps}
\\(c)
\caption{Distribution of axial velocity at different instants of time
  in the case of wave train propagation. When the pressure at the
  lower end of esophagus (LEE) is more than that at the upper end of
  esophagus (UEE), backward flow is enhanced, while the forward flow reduced. It is interesting to note that backward flow occurs at a faster rate even there is a small change of pressure at the LEE.}
\label{paper6_veloc6.2.9-6.2.15}
\end{center}
\end{figure}

  Figs. \ref{paper6_veloc6.1.9-6.1.15}-\ref{paper6_veloc6.2.9-6.2.15} present velocity distribution for situations where
  the pressure at the lower end of esophagus (LEE)
  is higher than that at the UEE (upper end of esophagus). When $\Delta$P rises, reflux region is extended
  for single bolus transport as well as train wave transport for all types of fluids including shear
  thickening case (although there is not significant pressure change due to increase in $P_1$).
  In addition, magnitude of velocity is reduced in the forward flow region, while in the reflux region it increases.

\begin{figure}
\includegraphics[width=7.2in,height=5.0in]{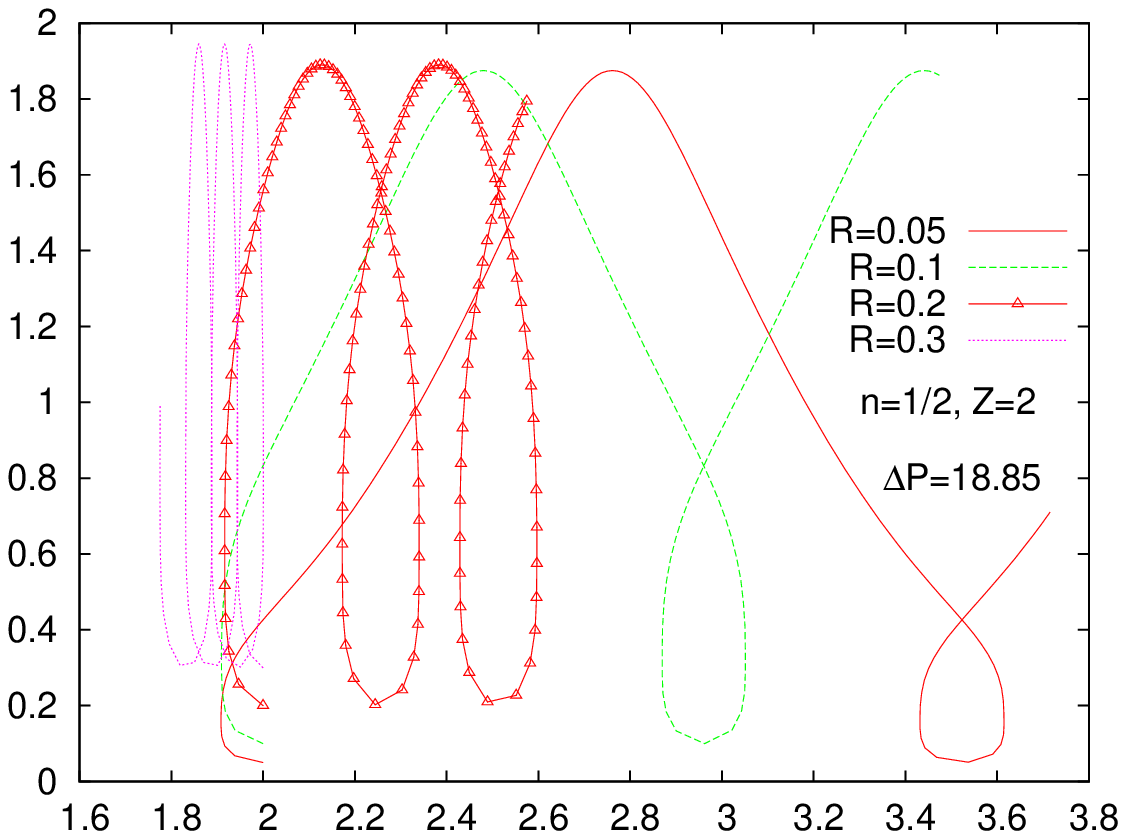}
\caption{Particle trajectories for peristaltic flow of a non-Newtonian
fluid of shear-thinning type at different locations (Z,R) (viz.
(2,0.05), (2,0.1), (2,0.2), (2,0.3)). Particles near the tube-axis travel
more distance in the axial direction and time taken by them to
complete the respective trajectories is less than their own particle
periods. Where as, particles near the boundary move less distance in
the axial direction and time taken to complete the trajectories is
more than the respective particle periods.}
\label{paper6_trajc6.1.eps}
\end{figure}

\subsection{Particle Trajectory and Reflux phenomenon}
\begin{figure}
\includegraphics[width=7.2in,height=5.0in]{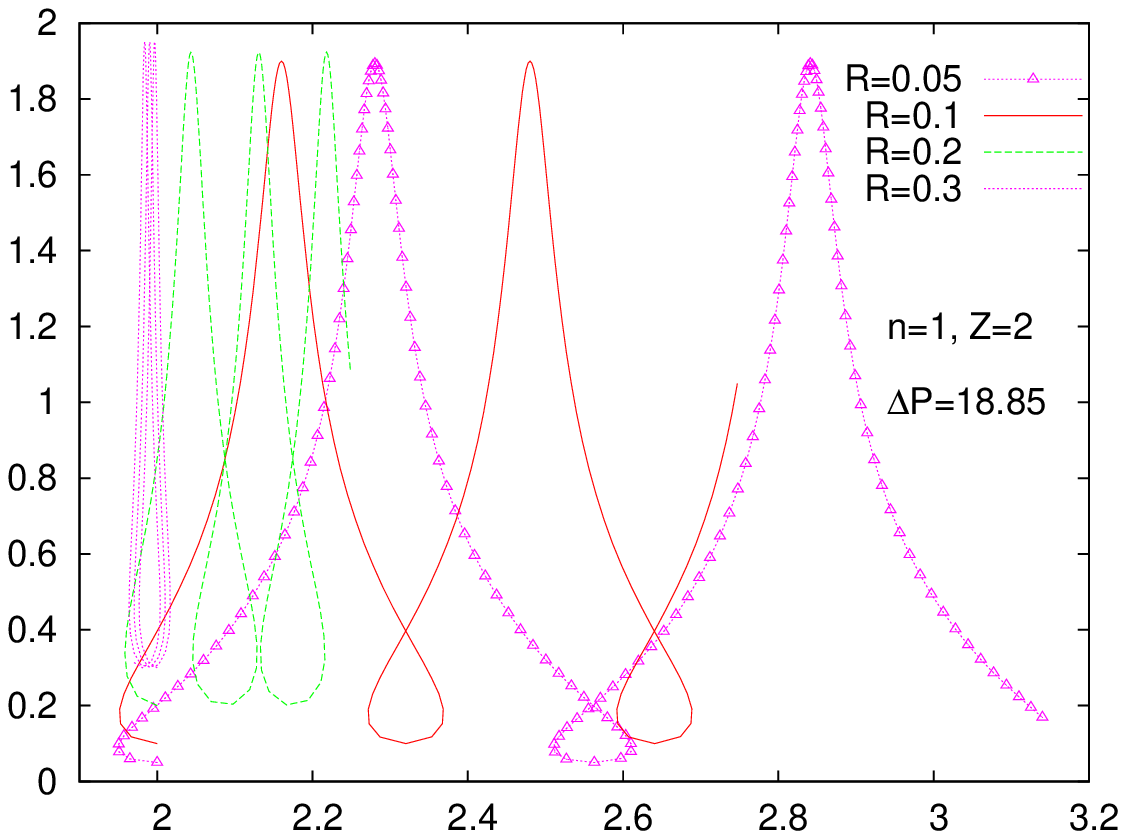}
\caption{Particle trajectory for peristaltic flow of a Newtonian fluid
  at different locations. Final position of the particle at
  (2,0.3) (located near the boundary) is (1.97,0.43). This particle
  slightly moves in the direction opposite to the wave.}
\label{paper6_trajc6.2.eps}
\end{figure}

\begin{figure}
\includegraphics[width=7.2in,height=5.0in]{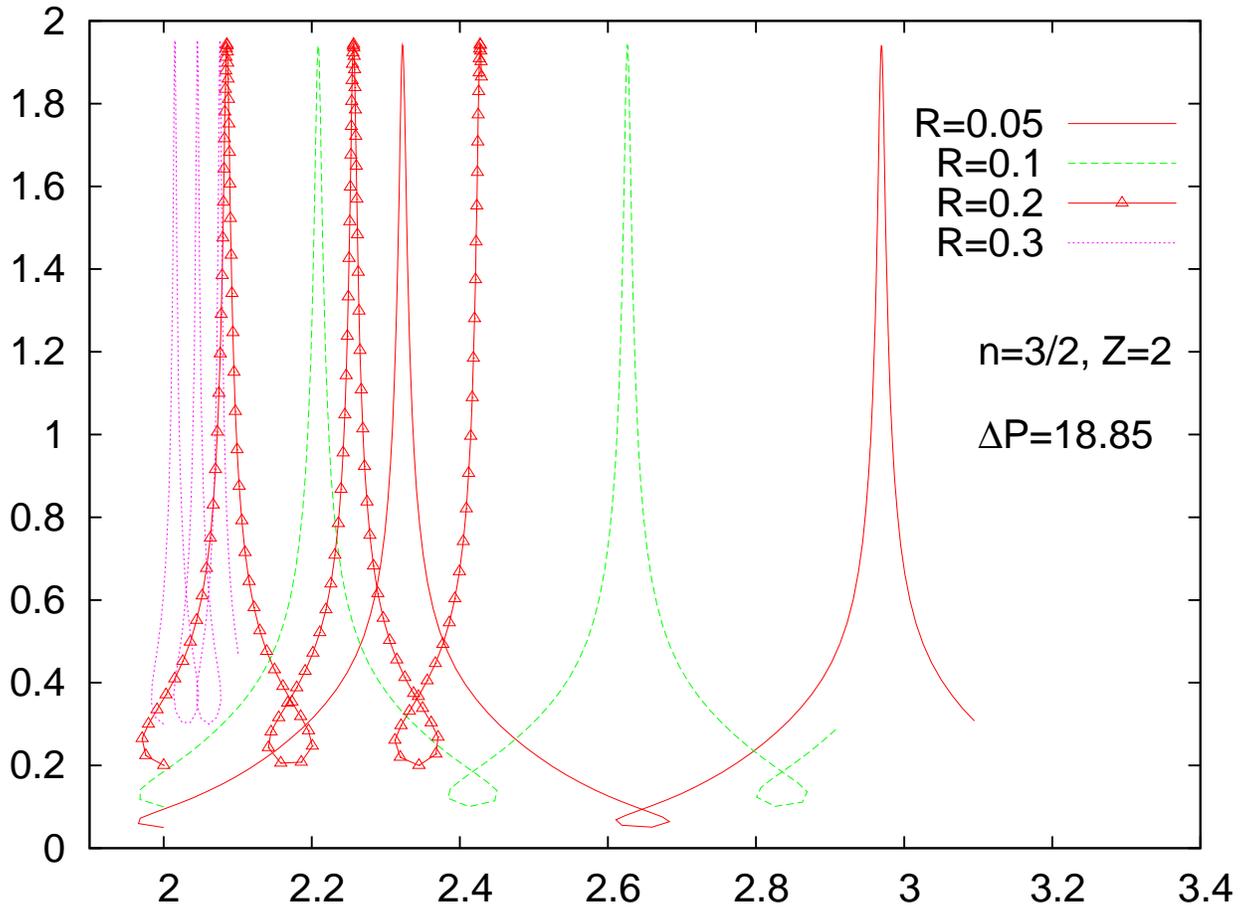}
\caption{Particle trajectories for peristaltic flow of a
  shear-thickening fluid at different locations. Final position of
  the particle at (2,0.3) (located near the boundary) is
  (2.1,0.47). All particles shown here move in direction same as that of the wave propagation.}
\label{paper6_trajc6.3.eps}
\end{figure}

It is known that one of the important characteristics of peristaltic
flow is the reflux phenomenon. It refers to the presence of fluid
particles that move in a direction opposite to that of the
peristaltic wave. In the infinite tube model reflux generally occurs
under conditions of partial occlusion and adverse pressure
difference across one wave length. The comparison of particle
trajectories corresponding to three different types of fluids in
esophagus are shown in Figs.
\ref{paper6_trajc6.1.eps}-\ref{paper6_trajc6.3.eps} for a wave train
propagating with sinusoidal shape, where non-integral number of
waves exists in the tube. To determine the trajectories of the
particles in the Lagrangian frame of reference, the simultaneous
differential equations

\begin{eqnarray}
\frac{dZ}{dt}=U,~~\frac{dR}{dt}=V
\end{eqnarray}
have been solved by using RK4 method successively, starting from the
initial location of the particles. The particles are initially taken
to be located in the vicinity of the most occluded point. The
results presented in Figs.
\ref{paper6_trajc6.1.eps}-\ref{paper6_trajc6.3.eps} have been computed
by taking the dimensionless pressure difference $\Delta P$ to be
18.85 (cf. \cite{Li1}). It may be noted that the particle
trajectories computed on the basis of the present study resemble
those presented by Li and Brasseur \cite{Li1}. It is also observed
that most particles in both Newtonian and rheological fluids undergo
a net positive displacement, while the particles nearest to the tube
wall move in the direction opposite to that of wave propagation.
Further, as rheological fluid index `n' increases, axial
displacement decreases and the particles reaching near the boundary
start moving slowly towards the axis at some points of time.
However, it is interesting to note that particles near the boundary
for a shear-thickening fluid move in the forward direction. Beyond
the most occluded region, it is observed that as the rheological
fluid index `n' increases, axial displacement of particles near the
axis increases.

\section{Concluding Remarks}
The motivation behind the present investigation is to study the
peristaltic transport of food bolus through the esophagus. For a
non-Newtonian fluid, local pressure is found to be very much
dependent on the fluid behaviour index `n'. The study shows that
variation in pressure (which is a local variable), forward and
retrograde flows and particle trajectories of the food bolus are all
highly sensitive to the length of the esophagus, the existence of
integral and non-integral number of waves in the tube as well as the
propagation of single/multiple waves in the esophagus. The leakage
of fluid is of common occurrence in the neighbourhood of an aortic
arch. The present study suggests that while designing a peristaltic
pump for all types of Newtonian/rheological fluids, it is quite
important to duly account for the unsteady effects arising out of
the variation in the length of the finite esophageal tube as well as
the differences between single
and multiple peristaltic wave propagation.

The present study is important so for as the movement of food
material through the esophagus is concerned, since due to shear
thinning effect, the viscosity of the fluid decreases with increase
in rate of shear stress. Tomato sauce/paste sauce, wheaped  cream
are some of the food materials that exhibit shear thinning effect.
Corn starch has been used as a non-Newtonian fluid in some
experimental studies. When water is mixed with cornstarch in a
certain proportion, the mixture, termed as Oobleck, possesses the
property of shear thickening.

When peristaltic waves start propagating, the circular muscle cells
shorten themselves and generate the contraction force. Involvement
of both the nerve control and intrinsic properties of muscle cells
makes the mechanism of muscle contraction somewhat complicated. The
peristaltic contraction is believed to act as an external force on
the tissue structure and travels downward with a certain speed.
Peristalsis in esophagus normally occurs by the propagation of a
single wave of active muscle contraction preceded by a single wave
of muscle relaxation. The motion of the wall is directly linked to
the pressures within the fluid (cf. \cite{Brasseur}) and there is a
relationship between the deformation of the esophageal wall as
recorded radiographically and the intra-bolus pressures as measured
manometrically during food bolus transport. Since the esophageal
wall is actively forced and the cavity volume at the contraction
region is forced to be reduced quickly in the contraction zone,
occlusion pressures are high. The bolus geometries in the
contraction region are associated with a rapid increase in pressure
toward the point of maximum occlusion. On the contrary, relaxation
is linked with a lack of muscle tone. The rate of local pressure is
so less when the fluid is shear thinning ($n<1$) and is very large
when the fluid is shear thickening ($n>1$) compared to a Newtonian
fluid. Hence the muscle contracts at a very slower rate when $n<1$,
while it contracts rapidly when $n>1$. Thus it is more comfortable
to swallow a food material having shear thinning properties than
food stuff possessing shear thickening characteristics. In the case
of a Newtonian fluid, the comfort in swallowing is not as easy as in
the case of shear thinning material, but is easier than in case of
shear thickening food material. Thus the study of the present
non-Newtonian model contributes to having a better understanding of
muscle movement. Moreover, the esophageal wall just distal to the
peristaltic wave must be passively forced open by the pressures
within the approaching food bolus. In order to overcome the thoracic
pressure exterior to the esophagus as well as any residual tension
within the esophageal wall, these intra-bolus pressure needs to be
sufficiently high. Figs. 2-5 reveal that at first the pressure rises
slowly and then increases rapidly to a peak as the contraction wave
passes, while relaxation is not related with rapid changes in
pressure.\\


{\bf Acknowledgment:} {\it The authors are highly thankful to both
the esteemed reviewers. The original manuscript could be
substantially revised on  the basis of their valuable comments. One
of the authors, S.Maiti is thankful to the Council of Scientific and
  Industrial Research (CSIR), New Delhi for awarding him an SRF.}

\end{document}